\documentclass[12pt]{article}
\def\timesbox{\hbox{$\scriptscriptstyle\times$}}
\def\ant{ {{\lower 1ex  \timesbox} \atop {\raise 1.5ex  \timesbox}}}
\newcommand\ZZZ{{\hbox{ Z\kern-1.6mm Z}}}
\newcommand{\beq}{\begin{equation}}
\newcommand{\eeq}{\end{equation}}
\newcommand{\bea}{\begin{eqnarray}}
\newcommand{\eea}{\end{eqnarray}}
\newcommand{\ra}{\rangle}
\newcommand{\la}{\langle}
\newcommand{\lt}{\left}
\newcommand{\rt}{\right}
\newcommand{\Iop}{\relax{\rm I\kern-.18em I}}
\newcommand{\Lop}{\relax{\rm I\kern-.18em L}}
\newcommand{\dop}{\relax{\rm I\kern-.8em d}}
\newcommand{\one}{{\hbox{ 1\kern-1.2mm l}}}

\newcommand{\dt}{\delta}

\newcommand{\xh}{\hat x}
\newcommand{\ph}{\hat p}
\newcommand{\pih}{\hat \pi}

\newcommand{\del}{\partial}

\newcommand{\Del}{\Delta}

\newcommand{\s}{\sigma}

\begin{document}

{}~
{}~
\hfill\vbox{\hbox{IMSc/2009/4/1}}
%\hbox{hep-th/yymmnnn}}
\break

\vskip 2cm

\centerline{\Large \bf DeWitt-Virasoro construction in}
\centerline{\Large \bf tensor representations}
              
\medskip

\vspace*{4.0ex}

\centerline{\large \rm Partha Mukhopadhyay }

\vspace*{4.0ex}

\centerline{\large \it The Institute of Mathematical Sciences}
\centerline{\large \it C.I.T. Campus, Taramani}
\centerline{\large \it Chennai 600113, India}

\medskip

\centerline{E-mail: parthamu@imsc.res.in}

\vspace*{5.0ex}

\centerline{\bf Abstract}
\bigskip

We generalize the DeWitt-Virasoro (DWV) construction of arXiv:0912.3987 [hep-th] 
to tensor representations of higher ranks. A rank-$n$ tensor state, which is by itself coordinate invariant, is expanded in terms of position eigenstates that transform as tensors of the same rank. The representation of the momentum operator 
in these basis states is then obtained by generalizing DeWitt's argument in Phys.Rev.85:653-661,1952. Such a representation is written in terms of certain bi-vector of parallel displacement and its covariant derivatives. With this machinery at hand we find tensor representations of the DWV generators defined in the previous work. The results differ from those in spin-zero representation by additional terms involving the spin connection. However, we show that the DWV algebra found earlier as a scalar expectation value remains the same, as required by consistency, as all the additional contributions conspire to cancel in various ways. In particular, vanishing of the anomaly term requires the same condition of Ricci-flatness for the background. 

\newpage

\tableofcontents

\baselineskip=18pt

\section{Introduction and summary}
\label{s:intro}

With the motivation of understanding exact conformal invariance of
string worldsheet theory in certain pp-wave background in hamiltonian
framework \cite{pp-wave} a particular approach to study the latter was
considered in \cite{pm0912}. The simplest case of a bosonic string
moving in an arbitrary metric-background was considered. The basic
idea was to find a way to describe the string quantum mechanics in a
coordinate independent way following the same line of argument that
was used by DeWitt \cite{dewitt52, dewitt57} for particle\footnote{Two-dimensional non-linear sigma model has been studied in lagrangian framework in \cite{lagrangian}. See \cite{hamiltonian} for earlier studies using hamiltonian framework. DeWitt's argument was used earlier in string theory in \cite{lawrence95}.}.   

The construction of \cite{pm0912}, hereafter called the DeWitt-Virasoro (DWV) construction, goes as follows. We first rewrite the worldsheet theory in terms of the Fourier modes in such a way that it can be interpreted as a single particle moving in an infinite-dimensional curved background subject to certain potential. Since the chosen degrees of freedom are non-local on the worldsheet the classical conformal invariance is not directly visible. However, the description possesses some more structure which look unusual from the point of view of a particle dynamics, but captures the underlying worldsheet. This enables us to write down the classical Virasoro generators in terms of the phase-space variables of the infinite-dimensional/particle description. Then implementing DeWitt's argument \cite{dewitt52, dewitt57} naturally leads to a quantum background independent definition for such generators, hereafter called DeWitt-Virasoro generators. The machinery available through \cite{dewitt52} enables one to calculate the matrix element of an arbitrary operator constructed out of the DWV generators between any two scalar states. The result written in position space is manifestly generally covariant. Using this tool we computed the DWV algebra in spin-zero representation. The result is given by the Witt algebra with certain additional anomalous terms that vanish for Ricci-flat backgrounds. 

In this work we will generalize the above construction to higher rank tensor representations in a certain sense. Before we delve into the details of this generalization we first try to motivate and explain what we intend to achieve.
String theory contains infinite number of higher rank tensor states. Around flat background such states are constructed by applying suitable creation operators on the ground state (which can be solved exactly). State-operator mapping of this sort should hold for any CFT representing an exact classical background in string theory and the set of all possible excited states, with suitable restriction, should correspond to the fluctuation of the tensor fields around that particular background\footnote{Background independence restricted to conformal backgrounds has been studied in the context of classical closed string field theory (CSFT) \cite{zwiebach92} in \cite{sen90, sen93}. Construction of CSFT for a non-conformal background was discussed in \cite{zwiebach96}.}. However, as we will argue, state-operator mapping, at least in this form, is not applicable in our generally covariant description. What appears to be more relevant in this case are the Schr$\ddot{\rm o}$dinger wavefunctions of the tensor states in position representation. In an arbitrary background these wavefunctions should correspond to the fluctuations of the tensor fields in spacetime. One would like to properly formulate this connection in a precise way. For example, one would like to understand how such wavefunctions should be explicitly constructed in order to be able to define the {\it string Hilbert space}. 

Before we attempt to answer these questions, in this work, we would like to understand certain computations that are entirely guided by the tensorial property of the wavefunctions and therefore can be carried out without knowing the explicit construction of those.  More precisely, our goal is to understand how to compute coordinate invariant matrix elements involving arbitrary tensor wavefunctions. To this end we describe a framework where the tensor wavefunctions are encoded into the states in a special way. This is done by introducing the position eigenstates which themselves are
higher rank tensors. The resulting analysis can be viewed as a tensor generalization of the work in \cite{dewitt52}. In that work the momentum operator in
position space was obtained by demanding that the canonical commutation
relations and the hermiticity conditions hold true as expectation
values between two position eigenstates of spin zero. The basic
ingredient that goes into such an analysis is the orthonormality and
completeness conditions for such basis states. A natural
generalization of the orthonormality condition for higher rank basis
states is given in terms of  bi-vector of parallel displacement.
A bi-vector is a bi-local object which carries a vector index
associated with each of the two points in the argument. Given two such points in
spacetime the parallel displacement bi-vector, contracted with another
vector at one of the points, gives the same vector parallel
transported to the other point along a given path. An example is the
bi-vector of geodetic parallel transport  as discussed in
\cite{dewitt60} where the path is the unique geodesic passing through
the two points. Our analysis is sensitive to the expansion of the
bi-vector only to first order in separation and therefore is independent of the
choice of path. It turns out that the desired representation of
the momentum operator can be written in terms of such bi-vectors and
their covariant derivatives. We use this to find the representation of the DWV generators as defined in our previous work. These expressions contain all the terms that are present in the scalar representation. Moreover, they receive additional contributions involving the spin connection. These are calculated by explicitly writing the parallel displacement bi-vector in terms of the vielbeins.

The DWV algebra was computed in \cite{pm0912} in spin-zero representation. Given that the calculation was done between two arbitrary scalar states, one expects the result to be true as operator equations and therefore is independent of the representation considered\footnote{I thank Romesh K. Kaul for emphasizing this point.}. Since in this work we have found the tensor representation from first principle, it is a good idea to check explicitly if this is true. We do this following the same basic method as described in \cite{pm0912}. As expected, the computation gets more complicated than the previous one because of the additional contributions. There arise three kinds of additional terms which depend on Riemann tensor, covariant derivative of spin connection and higher power of spin connection. It turns out that these terms are organized in such a way that they cancel each other due to various identities in general relativity reproducing indeed the same result for the DWV algebra.

The rest of the paper is organized as follows. The tensor representation has been constructed in section \ref{s:tensor-rep}. We discuss in what sense the existing framework in \cite{dewitt52, pm0912} is inadequate for the present purpose in \ref{ss:scope}. The new framework is described in \ref{ss:new}. Tensor representation of the momentum operator has been discussed in \ref{ss:momentum}. We use the results of section \ref{s:tensor-rep} to calculate the DWV generators and algebra in section \ref{s:DWV}. Various details of our technical derivations have been given in several appendices.

\section{Construction of tensor representations}
\label{s:tensor-rep}

\subsection{Scope of the existing framework}
\label{ss:scope}

As mentioned before, the analysis of \cite{pm0912} was done using an
infinite-dimensional language. We have summarized the relevant details of it in appendix \ref{a:summary}. All our discussion in this work will be done using the same language.

Below we will establish that the scope of the existing framework in \cite{dewitt52, pm0912} is as follows,
\bea
&& \hbox{\it Non-trivial invariant matrix elements can be
constructed only in scalar} \cr 
&& \hbox{\it representation.}
\label{scope}
\eea
Therefore, a new framework needs to be developed in order to construct the desired tensor representations.

We first discuss the kind of questions that are relevant for our present purpose. We define a tensor state to be one whose wavefunction is a tensor in the infinite-dimensional sense. More precisely, the latter can be obtained either by multiplying a tensor field with a scalar wavefunction or by applying covariant derivatives on it.  Given all possible tensor states, one would next like to understand how to compute the general coordinate invariant matrix elements of the DWV generators between such states. In position space such a matrix element should be given by an invariant integral where the integrand is constructed out of the tensor wavefunctions defined above.

Recall that in flat space the vacuum state can be explicitly solved and the tensor states are obtained by applying suitable creation operators on the vacuum state\footnote{Notice that the Schr$\ddot{\rm o}$dinger wavefunctions \cite{polchinski} of all such states are of the type that we are considering here.
As mentioned in section \ref{s:intro}, the aim of this work is not to find a background independent generalization, if it exists, of this procedure. Rather here we want to understand certain computations that are entirely guided by the tensorial properties of the wavefunctions.}. This is expected from the point of view of ordinary quantum mechanics. However, as we will demonstrate with an example below, this procedure is not suitable for our generally covariant construction.   

Using the existing framework of \cite{dewitt52, pm0912} one can calculate invariant matrix elements of the following form,
\bea
\la \chi|\hat{\cal O}|\psi \ra~,
\label{scalar-element}
\eea
where $|\chi\ra$ and $|\psi\ra$ are two arbitrary scalar states and $\hat {\cal O}$ is a scalar operator which can in general depend on momentum. For example, the scalar expectation value of the DWV algebra as computed in \cite{pm0912} is of the above form. Let us now consider the following two vector wavefunctions,
\bea
\chi^i(x) = a^i(x) \chi(x)~, \quad \psi^i(x) = a^i(x) \psi(x)~,
\eea
where $\chi(x)$ and $\psi(x)$ are the wavefunctions (see appendix \ref{a:summary} for our notations) of the states in (\ref{scalar-element}) respectively.
Usually the corresponding states will be defined as follows,
\bea
|\chi^i \ra = a^i(\hat x) |\chi\ra = \int dw~ \chi^i(x) |x\ra ~, \quad
|\psi^i\ra = a^i(\hat x) |\psi\ra = \int dw~\psi^i(x) |x\ra ~,
\label{GCT-O-element}
\eea
respectively. However, notice that the matrix element $\la \chi_i|\hat {\cal O}|\psi^i\ra$ is not coordinate invariant unless $\hat{\cal O}$ is a scalar field independent of momentum. This is simply because under a general coordinate transformation (GCT) this matrix element transforms as\footnote{GCT has been introduced in our quantum theory in appendix \ref{a:summary} below eq.(\ref{delta}). In appendix \ref{a:unitary} we have discussed how GCT is realized as a unitary transformation in tensor representations.},
\bea
\la \chi_i|\hat {\cal O}|\psi^i\ra \to
\la \chi_{j}|\lambda_i^{~j}(\hat x)\hat {\cal O}\lambda^i_{~k}(\hat x) |\psi^k\ra~.
\label{non-inv-matrix-element}
\eea
Since $\lambda_i^{~j}(\hat x)$ does not commute with $\hat{\cal O}$ if the latter depends on momentum, the 
above matrix element is not invariant in general. Proceeding in the similar
way one can argue that the only non-trivial (i.e. involving momentum operator)
invariant matrix elements that can be constructed are of the type
(\ref{scalar-element}). This establishes (\ref{scope}).

\subsection{A new framework}
\label{ss:new}

Given the above discussion, here we introduce a new framework where a tensor state is not created by applying operators on a scalar state, rather it is related to the corresponding wavefunction in a special way. The resulting construction will enable us to compute invariant matrix elements in tensor representations. 

In this framework a rank-$n$ tensor state $|\psi_{(n)}\ra$ is by itself GCT invariant. It is expanded in position space basis in the following way,
\bea
|\psi_{(n)}\ra = \int dw~ \psi^{i_1 i_2\cdots i_n}(x) |{}_{i_1i_2\cdots i_n} ; x\ra ~,
\label{psi-n}
\eea
where $\psi^{i_1 i_2\cdots i_n}(x)$ is the corresponding rank-$n$ tensor
wavefunction which is constructed following the general definition given in
the previous section. We have introduced the rank-$n$ position eigenstates $|{}_{i_1\cdots i_n}; x\ra$ satisfying the same eigenvalue equation as in (\ref{x-eigen-equation}). Under a GCT $\psi^{i_1 i_2\cdots i_n}(x)$ and $|{}_{i_1i_2\cdots i_n} ; x\ra$ transform as a contravariant and a covariant tensor of rank $n$ respectively (see appendix \ref{a:unitary}). The orthonormality and completeness conditions of the basis states are given by,
\bea
&&\la {}_{i_1\cdots i_n};x|{}_{j_1\cdots j_{\tilde n}};\tilde x\ra = \dt_{n,\tilde n}
\Del_{i_1\cdots i_n j_1\cdots j_n}(x,\tilde x) \dt(x,\tilde 
x)~,
\label{ortho} \\
&& \int dw~  |{}_{i_1\cdots i_n};x \ra g^{i_1j_1}(x) \cdots g^{i_n
  j_n}(x) \la {}_{j_1\cdots j_n};x| = 1~,
\label{compl}
\eea
where,
\bea
\Del_{i_1\cdots i_n j_1\cdots j_n}(x,\tilde x) &=& \Del_{i_1j_1}(x,\tilde x)
\cdots \Del_{i_nj_n}(x,\tilde x) ~,
\label{Del-n-def}
\eea
Equations (\ref{ortho}, \ref{compl}) are the tensor generalizations of eqs.(\ref{orthocomplete-0}). Before discussing what $\Del_{ij}(x,\tilde x)$ is we will first introduce
a bi-tensor in the sense of \cite{dewitt60} in our notations. A
bi-tensor $t_{i_1\cdots i_n j_1\cdots j_{\tilde n}}(x,\tilde x)$ is a
tensorial object with two sets of indices $\{i_1, \cdots, i_n\}$ and
$\{j_1,\cdots, j_{\tilde n}\}$ associated with the points $x$ and $\tilde x$
respectively such that it transforms as a tensor of rank $n$ and
$\tilde n$ respectively at the two points separately,
\bea
t_{i_1\cdots i_n j_1\cdots j_{\tilde n}}(x,\tilde x) \to t'_{i_1\cdots i_n
  j_1\cdots j_{\tilde n}}(x',\tilde x') &=& \lambda_{i_1}^{~k_1}(x) \cdots
\lambda_{i_n}^{~k_n}(x) \lambda_{j_1}^{~l_1}(\tilde x) \cdots
\lambda_{j_n}^{~l_{\tilde n}}(\tilde x) t_{k_1\cdots k_n l_1\cdots
  l_{\tilde n}}(x,\tilde x)~. \cr &&
\label{bi-tensor-trans}
\eea
In this sense $\dt(x,\tilde x)$ is a bi-scalar and $\Del_{ij}(x,\tilde x)$ is a
bi-vector. The latter is called a bi-vector of parallel
displacement which we have defined and explained the
properties of in appendix \ref{a:bivector}.

Certain comments regarding (\ref{ortho}) are in order. From the discussion of appendix \ref{a:bivector} it is
clear that in the coincidence limit the bi-vector approaches the metric,
\bea
\lim_{x\to \tilde x} \Del_{ij}(x,\tilde x) \to g_{ij}(\tilde x)~.
\label{Del-coincide}
\eea
Therefore it might appear unnecessary to use it in (\ref{ortho}) due to the presence of the delta function. The reason for this is as follows. First of all, notice that the expression in (\ref{ortho}) makes it manifest that the relevant matrix element is a bi-tensor. Secondly, the tensor representation of the momentum operator discussed in the next section will involve derivatives. As we will argue, such a quantity will be sensitive to the expansion of the bi-vector up to first order in separation.

Since we are using a complex basis, the hermitian conjugate of a basis state is
given by,
\bea
(|{}_{i_1, \cdots i_n}; x\ra)^{\dagger} = \la {}_{\bar i_1,\cdots \bar i_n}; x|~,
\label{ket-dagger}
\eea
and that of any operator ${\cal O}$ is defined to be\footnote{For
  example, for ${\cal O}=x^i$, using (\ref{hermiticity}) on the left
hand side of (\ref{Odagger}) one gets, $x^{\bar i} \dt_{n,\tilde n}
\Del_{i_1\cdots i_n j_1\cdots j_n}(x,\tilde x) \dt(x,\tilde x)$. Then using the reality property in (\ref{x-eigen-equation}) and,
\bea
\Del_{ij}(x,\tilde x)^* = \Del_{\bar i \bar j}(x,\tilde x)~,
\label{reality-Del}
\eea
one gets the same result from the right hand side of (\ref{Odagger}).},
\bea
\la {}_{i_1\cdots i_n};x| {\cal O}^{\dagger} |{}_{j_1\cdots
  j_{\tilde n}};\tilde x\ra = \la {}_{\bar j_1\cdots \bar j_{\tilde n}};\tilde x| 
{\cal O}
|{}_{\bar i_1\cdots \bar i_n};x\ra^*~.
\label{Odagger}
\eea
This completes the description of the present framework using which all the calculations of this paper will be done.

\subsection{Tensor representation of momentum operator}
\label{ss:momentum}

Here we will discuss the rank-$n$ tensor representation of the
momentum operator in position space. We will actually directly work
with the shifted momentum operators as defined in (\ref{pih-def}) \cite{omote72}.
The derivation of the representation of $\pih_k$ has been given in appendix
\ref{a:pi-rep}. The result is given by (in $\hbar = \alpha' =1$ unit),
\bea
\la{}_{i_1\cdots i_n};x|\pih_k|{}_{j_1\cdots j_{\tilde n}};\tilde x\ra
&=& -i \dt_{n,\tilde n} \lt[\Del_{i_1\cdots i_n j_1\cdots j_n}(x,\tilde x)
\del_k\dt(x,\tilde x) \rt. \cr
&& \lt. + F_{i_1\cdots i_n j_1\cdots j_n k}(x,\tilde x) \dt(x,\tilde x)\rt] ~, \cr
&=& -i \dt_{n,\tilde n} \lt[-\Del_{i_1\cdots i_n j_1\cdots j_n}(x,\tilde x)
(\tilde \del_k +\gamma_k(x)) \dt(x,\tilde x) \rt. \cr
&& \lt. + F_{i_1\cdots i_n j_1\cdots j_n k}(x,\tilde x) \dt(x,\tilde x)\rt] ~,
\label{pi-tensor-rep-ele}
\eea
where,
\bea
F_{i_1\cdots i_n j_1\cdots j_n k}(x,\tilde x) &=& - \nabla_k \Del_{i_1\cdots  i_n j_1\cdots j_n}(x,\tilde x) ~.
\label{F-n-def}
\eea
where $\nabla_k$ is the covariant derivative.
We first notice that the above expressions display the expected
tensorial properties of the relevant matrix element. The two equations in
(\ref{pi-tensor-rep-ele}) are related by the second identity in (\ref{del-identities}).
Because of the presence of delta functions in
(\ref{pi-tensor-rep-ele}) the complete functional dependence of
$\Del_{i_1\cdots i_n j_1\cdots j_n}(x,\tilde x)$ and $F_{i_1\cdots i_n
  j_1\cdots j_n k}(x,\tilde x)$ are not needed. The following expressions,
which can be obtained (see appendix \ref{a:pi-rep}) by expanding these
quantities to the desired order, will be used for all practical purposes,
\bea
\la{}_{i_1\cdots i_n};x|\pih_k|{}_{j_1\cdots j_{\tilde n}};\tilde x\ra
&=& -i \dt_{n,\tilde n} \lt[g_{i_1\cdots i_n j_1\cdots
  j_n}(\tilde x)\del_k\dt(x,\tilde x) \rt. \cr && \lt. -\lt( \omega_{i_1\cdots
  i_n j_1\cdots j_n k}(x)+\gamma_{i_1\cdots i_n j_1\cdots j_n k}(x)
\rt) \dt(x,\tilde x) \rt] ~, \cr
&=& -i \dt_{n,\tilde n} \lt[-g_{i_1\cdots i_n j_1\cdots
  j_n}(x)(\tilde \del_k+\gamma_k(x))\dt(x,\tilde x) \rt.\cr &&
\lt. -\lt(\omega_{i_1\cdots i_n j_1\cdots j_n, k}(x)-\gamma_{j_1\cdots
  j_n i_1\cdots i_n k}(x) \rt) \dt(x,\tilde x)
\rt] ~.
\label{pi-tensor-rep}
\eea
Various tensors appearing in the above equations are given by,
\bea
g_{i_1\cdots i_n j_1\cdots j_n} &=& g_{i_1j_1}\cdots g_{i_nj_n}~,
\label{g-n-def}\\
\omega_{i_1\cdots i_n j_1\cdots j_n k} &=&
\omega_{ki_1j_1}g_{i_2j_2}\cdots g_{i_nj_n}+
g_{i_1j_1}\omega_{ki_2j_2}g_{i_3j_3}\cdots g_{i_nj_n} +\cdots ~,
\label{omega-n-def} \\
\omega_{kij} &=& \nabla_k e^{\hat i}_{~i} \eta_{\hat i \hat j}e^{\hat
  j}_{~j} = \omega_{k\hat i \hat j} e^{\hat i}_{~i} e^{\hat j}_{~j}~, \cr
&=& -\omega_{kji}~,
\label{omega-kij} \\
\gamma_{i_1\cdots i_n j_1\cdots j_n k} &=&
\gamma_{ki_1j_1}g_{i_2j_2}\cdots g_{i_nj_n}+
g_{i_1j_1}\gamma_{ki_2j_2}g_{i_3j_3}\cdots g_{i_nj_n} +\cdots ~, \cr
\gamma_{kij} &=& \gamma^{i'}_{ki}g_{i'j}~,
\label{gamma-n-def}
\eea
where $e^{\hat i}_{~i}$, $\eta_{\hat i \hat j}$ and $\omega_{k\hat i
  \hat j}$ are the infinite-dimensional analogues of the vielbein, the
Minkowski metric and the spin connection - see equations
(\ref{vielbein-inf}) and (\ref{omega}).

Notice that in the first terms of both the expressions in
(\ref{pi-tensor-rep}) the argument of $g_{i_1\cdots i_n j_1\cdots
  j_n}$ and the variable with respect to which the derivative applies
on the delta function are different. This is how we choose the
argument to expand $\Del_{i_1\cdots i_n j_1\cdots j_n}(x,\tilde x)$
around. Because of this choice the derivative can pass through the
factor $g_{i_1\cdots i_n j_1\cdots j_n}$ freely. This form is useful
for performing various manipulations that will be used extensively in
our calculation of the DWV algebra.

\section{DWV construction}
\label{s:DWV}

The right and left moving DWV generators $\hat L_{(i)}$ and $\hat{\tilde
  L}_{(i)}$ respectively are defined in the same way as done in \cite{pm0912},
\bea
4 \hat L_{(i)} = \hat K_{(i)}-\hat Z_{(i)}+\hat V_{(i)} ~, \quad
4 \hat{\tilde L}_{(i)} = \hat K_{(\bar i)}+\hat Z_{(\bar i)}+\hat V_{(\bar i)}~,
\label{LLtilde}
\eea
where,
\bea
\hat K_{(i)} &=& \pih^{\star}_k g^{k l+i}(\xh) \pih_l~, \cr
\hat Z_{(i)} &=& \hat Z^L_{(i)} + \hat Z^R_{(i)}~, \cr
\hat V_{(i)} &=& g_{kl}(\xh) a^k(\xh) a^{l+i}(\xh)~,
\label{KZV-quantum}
\eea
where,
\bea
\hat Z^L_{(i)} = \pih^{\star}_k a^{k+i}(\xh)~, \quad \hat Z^R_{(i)} = a^{k+i}(\xh) \pih_k ~,
\label{ZLZR}
\eea
Notice that both $\hat Z^L_{(i)}$ and $\hat Z^R_{(i)}$ are coordinate
invariant and both have the right classical limit.  The relative
coefficient between these two terms in the second equation of
(\ref{KZV-quantum}) is fixed by the hermiticity property of the DWV generators,
\bea
\hat L_{(i)}^{\dagger} = \hat L_{(\bar i)} ~, \quad \hat{\tilde
  L}_{(i)}^{\dagger} = \hat{\tilde L}_{(\bar i)} ~.
\eea
Given eqs.(\ref{pi-tensor-rep}), one can compute the matrix elements
of $\hat K_{(i)}$, $\hat Z^L_{(i)}$ and $\hat Z^R_{(i)}$ between two
arbitrary tensor states. The manipulations involved in such
calculations have been demonstrated in deriving
eq.(\ref{manipulation}). The final results are given by\footnote{Notice that according to
  eqs.(\ref{pi-tensor-rep-ele}) $\pi_k$ connects tensor states
  of the same rank only.},
\bea
\la\chi_{(n)}|\hat K_{(i)}|\psi_{(n)}\ra
&=& (-i)^2 \int dw~ \lt[\bar \chi^{i_1\cdots i_n} \nabla_{(i)}^2
\psi_{i_1\cdots i_n} + \nabla^{k+i} \bar \chi^{i_1\cdots
  i_n}\omega_{i_1\cdots i_n j_1\cdots j_n k}\psi^{j_1\cdots j_n}
\rt. \cr
&& - \bar \chi^{i_1\cdots i_n}\omega_{i_1\cdots i_n j_1\cdots
  j_n k}\nabla^{k+i}\psi^{j_1\cdots j_n} \cr
&& \lt. + \bar \chi^{i_1\cdots
  i_n}\omega_{i_1\cdots i_n k_1\cdots k_n k}\omega^{k_1\cdots
  k_n~~~~~k+i}_{~~~~~~j_1\cdots j_n}\psi^{j_1\cdots j_n}\rt]~,
\label{chi-K-psi} \\
\la\chi_{(n)}|\hat Z^L_{(i)}|\psi_{(n)}\ra &=& i \int dw~
\lt(\nabla_k\bar \chi_{i_1\cdots i_n} + \bar \chi^{j_1\cdots j_n}\omega_{j_1\cdots j_n i_1\cdots i_n k} \rt)
a^{k+i}\psi^{i_1\cdots i_n}~,
\label{chi-ZL-psi} \\
\la\chi_{(n)}|\hat Z^R_{(i)}|\psi_{(n)}\ra &=& -i \int dw~ \bar \chi^{i_1\cdots i_n}a^{k+i} (\nabla_k\psi_{i_1\cdots i_n}-
\omega_{i_1\cdots i_n j_1\cdots j_n k}\psi^{j_1\cdots j_n} )~,
\label{chi-ZR-psi}
\eea
where $\bar \chi^{i_1\cdots i_n} = \chi^{*\bar i_1 \cdots \bar i_n}$ and $\nabla_{(i)}^2 =\nabla_k\nabla^{k+i}$ is the {\it shifted Laplace operator} \cite{pm0912}. Notice that only the first terms in the above expressions were found in \cite{pm0912} for scalar representation. The spin connection terms, which originate from the $F$-term in (\ref{pi-tensor-rep-ele}), are the additional contributions due to tensor generalization.

In \cite{pm0912} the DWV algebra was calculated as expectation value between two spin-zero states. The result, written in operator form, is given by,
\bea
[\hat L_{(i)}, \hat L_{(j)}] &=& (i-j)  \hat L_{(i+j)} + \hat A^R_{(i)(j)}~, \cr
[\hat{\tilde L}_{(i)}, \hat{\tilde L}_{(j)}] &=& (i-j) 
\hat{\tilde L}_{(i+j)} + \hat A^L_{(i)(j)}~, \cr
[\hat L_{(i)}, \hat{\tilde L}_{(j)}] &=& \hat A_{(i)(j)}~.
\label{DWValg}
\eea
The anomaly terms are given by,
\bea
\hat A^R_{(i)(j)} &=& 0~, \cr
\hat A^L_{(i)(j)} &=& 0~, \cr
\hat A_{(i)(j)} &=& {1\over 8} \lt(\hat \pi^{\star k+i}
r_{kl}(\hat x) a^{l+\bar j}(\hat x) - a^{k+i}(\hat x) r_{kl}(\hat x) \hat \pi^{l+\bar j} \rt)~,
\label{DWVanomaly}
\eea
where $r_{ij}(x)$ is the Ricci  tensor in the infinite-dimensional
spacetime which, according to the general map (\ref{th-rule}),
is related to the same in physical spacetime namely, $R_{\mu \nu}(X)$
in the following way,
\bea
r_{ij} (x) \sim 2\pi \dt(0) \oint {d\s\over 2\pi} ~R_{\mu \nu}(X(\s)) e^{i(m+n)\s}~.
\label{Ricci}
\eea

The DWV algebra, as operator equations, should be valid independent of the representation considered. Therefore, the result in (\ref{DWValg}) is correct upto possible additional operator terms whose expectation value vanishes in the calculation of \cite{pm0912}. However, since the calculation was done between two arbitrary states, it is expected that such terms are not present. 

Notice that we have found the tensor representations in the previous section from first principle. This construction does not {\it a priori} know about the DWV algebra. Consistency would therefore require the results in (\ref{DWValg}, \ref{DWVanomaly}) to hold true as expectation value between two arbitrary tensor states of any rank. Following the basic method of \cite{pm0912} we have checked this consistency condition to affirmative in appendix \ref{a:alg}.

\begin{center}
{\bf Acknowledgement}
\end{center}

I am thankful to A. P. Balachandran, G. Date, T. R. Govindarajan, R. K. Kaul and B. Sathiapalan for illuminating discussions.

\appendix

\section{Summary of previous work}
\label{a:summary}

All the analysis in this paper has been done using the
infinite-dimensional/particle language introduced in
\cite{pm0912}. Here we will briefly review the basic results of that
work. The matter part of the conformally gauge fixed worldsheet
lagrangian takes the following form in this language,
\bea
L(x, \dot x) &=& {1\over 2} g_{ij}(x) \lt[\dot x^i \dot x^j -  a^i(x) a^j(x) \rt]~,
\label{Linfinite}
\eea
where the infinite-dimensional spacetime index is give by,
\bea
i = \{\mu, m \}~,
\label{i-mu-m}
\eea
$\mu = 0\cdots D-1$ and $m \in Z$ being the physical spacetime index
and the string-mode-number respectively\footnote{Throughout
the paper we make the following type of index identifications: $i=\{\mu,
m \}$, $j=\{\nu, n \}$, $k=\{\kappa, q\}$. \label{index}}. The
coordinate $x^i$ is implicitly dependent on the worldsheet time
coordinate $\tau$. A dot indicates derivative with respect to $\tau$. The map
between the  infinite-dimensional and worldsheet languages is given by,
\bea
x^i &=& \oint {d\s \over 2\pi}~ X^{\mu}(\s) e^{-im\s} ~,\cr
g_{ij}(x) &=& \oint {d\s \over 2\pi} ~G_{\mu \nu}(X(\s))e^{i(m+n)\s}~, \cr
a^i(x) &=&  \oint {d\s \over 2\pi}~ \del X^{\mu}(\s) e^{-im\s}~,
\label{xga-XGdX}
\eea
where $G_{\mu \nu}$ is the metric in physical spacetime, $\oint
\equiv \int_0^{2\pi}$ and $\del \equiv \del_{\s}$. Using the above map
an arbitrary field in the infinite-dimensional spacetime can be mapped
to a non-local worldsheet operator. A class of examples, which
includes the classical Virasoro generators and their quantum version, is given by a
multi-indexed object $u^{i_1j_1\cdots}_{i_2j_2\cdots}(x)$ constructed
out of the metric, its inverse, their derivatives and $a^i(x)$ (but
not its derivatives) such that $u^{i_1j_1\cdots}_{i_2j_2\cdots}(x)$
can not be factored into pieces which are not contracted
with each other. In this case one can construct a local worldsheet operator
 $U^{\mu_1 \nu_1 \cdots}_{\mu_2 \nu_2 \cdots}(X(\s))$ simply
by performing the following replacements in the expression of
$u^{i_1j_1\cdots}_{i_2j_2\cdots}(x)$,
\bea
g_{ij}(x) \to G_{\mu \nu}(X(\s))~, \quad g^{ij}(x) \to G^{\mu
  \nu}(X(\s))~,\quad \del_i \to
\del_{\mu}~, \quad a^i(x) \to \del X^{\mu}(\s)~.
\label{replace}
\eea
The two objects $u^{i_1j_1\cdots}_{i_2j_2\cdots}(x)$ and $U^{\mu_1
\nu_1 \cdots}_{\mu_2 \nu_2  \cdots}(X(\s))$ are related to each other
by the following general rule,
\bea
u^{i_1j_1\cdots}_{i_2j_2\cdots}(x) \sim [2\pi \delta (0)]^N \oint {d\s \over 2\pi} ~U^{\mu_1
\nu_1 \cdots}_{\mu_2 \nu_2 \cdots}(X(\s)) e^{i(m_2+n_2+\cdots)\s -i(m_1+n_1+ \cdots)\s}~,
\label{th-rule}
\eea
where $N$ is the number of traces in $u$\footnote{For example, $\gamma_k$ as
  defined in (\ref{chr}) has a single trace and therefore is divergent:
\bea
\gamma_k(x) = 2\pi \dt(0) \oint {d\s \over 2\pi}~ \Gamma_{\kappa}(X(\s)) e^{iq\s}~,
\eea
where the expression for $\Gamma_{\kappa}(X)$ can be read out from (\ref{chr}) using the replacement
(\ref{replace}). The Ricci tensor is another example
of a single-trace field (see eq.(\ref{Ricci})).} and the argument of the Dirac delta function $\dt (0)$
appearing on the right hand side is the worldsheet space direction,
\bea
\dt(0) = \lim_{\s \to \s'} \dt (\s-\s') =\lim_{\s\to \s'} {1\over 2\pi} \sum_{n\in
Z}e^{in(\s-\s')}~.
\label{delta-0}
\eea
See \cite{pm0912} for further discussion on the appearance of such $\dt(0)$-factors. In the infinite-dimensional language the problem at hand possesses certain shift properties which can be written as,
\bea
u^{i_1+i i_2 \cdots}_{j_1j_2\cdots} &=& u^{i_1 i_2+i
  \cdots}_{j_1j_2\cdots} = u^{i_1 i_2 \cdots}_{j_1-i j_2\cdots} =
u^{i_1 i_2 \cdots}_{j_1j_2-i\cdots} = \cdots ~.
\label{shift}
\eea
A shift in the infinite-dimensional index is defined to be $i+j=\{\mu,
m+n\}$\footnote{Notice that we choose to denote the physical
  spacetime index corresponding to $i+j$ by the one associated with
  the first index (i.e. $i$) appearing in the shift. We will follow
  this convention in all our expressions.}. It is obvious that (\ref{shift}) is a
direct consequence of (\ref{th-rule}). Last equation in
(\ref{xga-XGdX}) implies,
\bea
\del_ja^k = i(j) \delta^k_j~.
\label{del-a}
\eea
where given the spacetime index $i$ as in footnote \ref{index}, we have
defined $(i)=m$. An important identity which was proved in \cite{pm0912} and will be
used crucially in deriving the results in appendix \ref{a:alg} is given by,
\bea
a^{k+i}\del_k u^{i_1 j_1\cdots}_{i_2j_2\cdots} &=&
i\{(i+i_1+j_1+\cdots ) - (i_2+j_2 +\cdots )\} u^{i_1+i j_1\cdots}_{i_2j_2\cdots}~.
\label{id}
\eea

To quantize the theory we first define the conjugate momentum $p_i
={\del L \over \del \dot x^i}$ and then impose (in $\hbar = \alpha' =1$ unit),
\bea
[\xh^i, \ph_j] = i \delta^i_j ~,
\label{can-comm}
\eea
where we have used a hat to denote a quantum operator. The hermiticity
properties are given by,
\bea
(\hat x^i)^{\dagger} = \hat x^{\bar i}~, \quad \hat p_i^{\dagger} =
\hat p_{\bar i}~,
\label{hermiticity}
\eea
where given (\ref{i-mu-m}) we have defined,
\bea
\bar i = \{\mu, -m\}~.
\eea

In \cite{dewitt52, dewitt57} it was shown how to construct the position space representation of this quantum theory such that general covariance is manifest. An arbitrary scalar state $|\psi \ra$\footnote{Throughout this work we will suppress the $\tau$ dependence of the Schr$\ddot{\rm o}$dinger states to reduce clutter.} is expanded in position basis in the following way,
\bea
|\psi\ra = \int dw~ \psi(x) |x\ra ~,
\label{psi-ket}
\eea
where $dw=dx \sqrt{g}$ ($g =|{\rm det}g_{ij}|$) is the invariant measure. The position eigenstate $|x\ra$ satisfies 
\bea
\hat x^i |x\ra = x^i |x\ra ~, \quad (x^i)^* = x^{\bar i}~,
\label{x-eigen-equation}
\eea
and the following orthonormality and completeness conditions,
\bea
\la x |\tilde x\ra = \dt(x, \tilde x)~, \quad \int dw ~|x\ra \la x| = 1~,
\label{orthocomplete-0}
\eea
where,
\bea
\dt(x,\tilde x) = \lt(g(x) g(\tilde x)\rt)^{-{1\over 4}} \dt(x-\tilde x) ~,
\label{delta}
\eea
$\dt(x-\tilde x)$ being the Dirac delta function.

The general coordinate transformation (GCT) is understood in the quantum theory in the following way.
The eigenvalue $x^i$ of the position operator is the coordinate of a point ${\cal P}$ on the manifold ${\cal M}$ where the particle-dynamics is taking place. A GCT is given by a set of functions $f=\{f^i(x)|f^i(x)^*=f^{\bar i}(x)\}$ such that,
\bea
&& \hat x^i \to \hat x'^i = f^i(\hat x)~, \quad \hat p_i \to \hat p'_i = {1\over 2} (\lambda_i^{~j}(\hat x) \hat p_j + \hat p_j \lambda_i^{~j}(\hat x))~,  \cr
&& x^i \to x'^i = f^i(x)~,
\label{GCT}
\eea
where,
\bea
\lambda^i_{~j}(x) = {\del x'^i \over \del x^j}~, \quad \lambda^{~j}_i(x) = {\del x^j \over \del x'^i}~,
\label{lambda}
\eea
are the Jacobian matrix for the transformation and its inverse respectively. The original system ($\hat x^i$, $\hat p_i$, $x^i$) and the transformed one ($\hat x'^i$, $\hat p'_i$, $x'^i$) give equivalent descriptions of the quantum theory such that $x^i$ and $x'^i$ are associated to the same point ${\cal P}$ in ${\cal M}$. Both the wavefunction and the position eigenstate in (\ref{psi-ket}) are scalar under GCT,
\bea
\psi(x) \to \psi'(x') = \psi(x)~, \quad |x\ra \to |x'\ra' = |x\ra~. 
\eea

The DWV generators are defined in (\ref{LLtilde}, \ref{KZV-quantum}, \ref{ZLZR}) in terms of certain shifted momentum operators,
\bea
\pih_k = \ph_k + {i \over 2} \gamma_k(\xh)~, \quad
\pih_k^{\star} = \ph_k - {i \over 2} \gamma_k(\xh)~,
\label{pih-def}
\eea
where $\gamma_k$ are the contracted Christoffel symbols,
\bea
\gamma_k = \gamma^i_{ki}~, \quad \gamma^i_{jk} = {1\over 2} g^{il}\lt(\del_j g_{lk} + \del_k
g_{lj} - \del_l g_{jk} \rt)~.
\label{chr}
\eea
Position space representation of $\hat \pi_k$ is found to be,
\bea
\la x |\hat \pi_k |\tilde x\ra = -i \del_k \dt(x,\tilde x)~.
\label{pi-rep-0}
\eea
Using either the definitions in (\ref{pih-def}) or the above representation one can establish the following transformations under GCT,
\bea
\hat \pi_k \to \hat \pi'_k = \lambda_k^{~l}(\hat x) \hat \pi_l~, \quad  \hat \pi^{\star}_k \to \hat \pi^{\star '}_k = \hat \pi^{\star}_l \lambda_k^{~l}(\hat x) ~,
\label{GCT-pi}
\eea
which show that the DWV generators are scalar operators. Using this basic ingredient one calculates the DWV algebra in spin-zero representation. The result is as given in eqs.(\ref{DWValg}, \ref{DWVanomaly}).

\section{Bi-vector of parallel displacement}
\label{a:bivector}

Here we will define and discuss the properties of the parallel
displacement bi-vector $\Del_{ij}(x,\tilde x)$. As explained in
section \ref{s:intro}, such a bi-vector contracted with another vector at
one of the points, gives the same vector parallel transported to the
other point along a given path. However, as remarked below
eq.(\ref{Del-coincide}), our analysis is sensitive to the expansion of
$\Del_{ij}(x,\tilde x)$ only to first order in $(x-\tilde x)^i$. Such
contributions are independent of the path chosen. To make the
discussion valid for arbitrary separation and for definiteness we choose to
discuss the bi-vector of geodetic parallel displacement
\cite{dewitt60} where the aforementioned path is a geodesic. We will
first define this object (following \cite{dewitt60}) for a particle in the
physical spacetime using our notation and elaborate, to some extend, on the
explicit representation in terms of the vielbeins. Then we will consider
the string-case and describe how it is constructed in the
infinite-dimensional language.

Let us consider two nearby points $X^{\mu}$ and $\tilde X^{\mu}$ in the physical
spacetime such that a unique geodesic passes through them. The
world-function $\Sigma(X,\tilde X)$, which is a bi-scalar, is given
by\cite{synge60}\footnote{The proportionality constant is usually
  taken as $\pm {1\over 2}$ with the $\pm$ sign corresponding to
  space-like and time-like separations respectively. However, for our purpose such
  details are not needed.},
\bea
\Sigma(X,\tilde X) &\propto & S(X,\tilde X)^2~,
\eea
where $S(X,\tilde X)=S(\tilde X,X)$ is the bi-scalar of geodetic interval.
The bi-vector of geodetic parallel displacement $D_{\mu \nu}(X,\tilde X)$
satisfies the following defining equations \cite{dewitt60},
\bea
\nabla^{\rho}\Sigma(X,\tilde X) \nabla_{\rho}D_{\mu \nu}(X,\tilde X) &=& 0~, \cr
\tilde \nabla^{\rho}\Sigma(X,\tilde X) \tilde \nabla_{\rho}D_{\mu \nu}(X,\tilde X) &=& 0~, \cr
\lim_{X\to \tilde X} D_{\mu \nu}(X,\tilde X) &=& G_{\mu \nu}(\tilde X)~,
\label{D-def}
\eea
where $\nabla_{\mu}$ and $\tilde \nabla_{\mu}$ denote the covariant
derivatives with respect to $X^{\mu}$ and $\tilde X^{\mu}$ respectively. It turns out that
$\nabla^{\rho}\Sigma(X,\tilde X)$ and $\tilde \nabla^{\rho}\Sigma(X,\tilde X)$ are
directed along (or opposite to) the tangent vectors to the geodesic at $X$ and $\tilde X$ respectively \cite{synge60}. From this one infers the following
geometrical interpretation of $D_{\mu \nu}(X,\tilde X)$,
\bea
D_{\mu}^{~\nu}(X,\tilde X) V_{\nu}(\tilde X) = V_{\mu}(\tilde X\to X)~, \quad D^{\mu}_{~\nu}(X,\tilde X) V_{\mu}(X) = V_{\nu}(X\to \tilde X)~,
\label{D-int}
\eea
where $V_{\mu}(X)$ is an arbitrary vector and $V_{\mu}(X\to \tilde X)$ denotes
the vector $V_{\mu}(X)$ parallel transported from $X$ to
$\tilde X$. Therefore $D_{\mu \nu}(X,\tilde X)$ parallel displaces a vector
along the geodesic. As argued in \cite{dewitt60}, $D_{\mu \nu}(X,\tilde X)$
is unique. Therefore eqs.(\ref{D-int}) can be taken as an alternative definition. The following properties follow from this geometrical interpretation,
\bea
D_{\mu \nu}(X,\tilde X) &=& D_{\nu \mu}(\tilde X,X) ~, \cr
D_{\mu \rho}(X,\tilde X) D_{\nu}^{~\rho}(X,\tilde X) &=& G_{\mu \nu}(X)~, \cr
D_{\rho \mu}(X,\tilde X) D_{~\nu}^{\rho}(X,\tilde X) &=& G_{\mu \nu}(\tilde X)~.
\label{D-prop}
\eea

For our calculations we will use the explicit representation of $D_{\mu \nu}(X,\tilde X)$ in terms of the vielbeins $E ^{\hat  \mu}_{~\mu}(X)$,  where $\hat \mu$ is the local Lorentz index. The representation is given by,
\bea
D_{\mu \nu}(X,\tilde X) &=& E^{\hat \mu}_{~\mu}(\tilde X\to X) \eta_{\hat \mu \hat \nu} E^{\hat \nu}_{~\nu}(\tilde X)~, \cr
&=& E^{\hat \mu}_{~\mu}(X) \eta_{\hat \mu \hat \nu} E^{\hat \nu}_{~\nu}(X\to \tilde X)~,
\label{D-rep}
\eea
where the parallel transport of the vielbeins is defined by vanishing
of the following covariant derivative,
\bea
\nabla_{\mu} E^{\hat \mu}_{~\nu} &\equiv & \del_{\mu} E^{\hat \mu}_{~\nu} -
\Gamma^{\kappa}_{\mu \nu}E^{\hat \mu}_{~\kappa} = - \Omega_{\mu ~~\hat
  \nu}^{~~\hat
  \mu} E^{\hat \nu}_{~\nu} ~,
\label{cov-der-vier}
\eea
along the geodesic, such that the Lorentz index remains unaffected. In
the last step above we have written the result in terms of the spin
connection $\Omega_{\mu}$, using the fact that the {\it total covariant
derivative} of the vielbein vanishes.

The simplest way to justify the representation (\ref{D-rep}) is to show that it leads to the same geometrical interpretation in (\ref{D-int}). According to (\ref{cov-der-vier}) the Lorentz component $V^{\hat \mu}= E^{\hat \mu}_{~\mu} V^{\mu}$ of a vector $V^{\mu}$ remains constant under parallel transport. Therefore, to parallel
transport a vector it is sufficient to transport only the vielbein,
\bea
V^{\mu}(\tilde X\to X) &=& E^{\hat \mu \mu}(\tilde X\to X) \eta_{\hat \mu \hat \nu} V^{\hat \nu}(\tilde X)~, \cr
&=& E^{\hat \mu \mu}(\tilde X\to X) \eta_{\hat \mu \hat \nu} E^{\hat \nu}_{~\nu}(\tilde X) V^{\nu}(\tilde X)~.
\eea
Combining the last equation above and the first equation in (\ref{D-int})
one identifies the first equation in (\ref{D-rep}). The second
equation in (\ref{D-int}) can also be obtained in a similar
way. These two equations make the first property in (\ref{D-prop})
manifestly true. It is also easy to prove the other two properties in
(\ref{D-prop}) by recalling the fact that a vielbein remains so under a parallel
transport \cite{synge60}. So we must have,
\bea
E^{\hat \mu}_{~\mu}(\tilde X\to X) E_{\hat \mu \nu}(\tilde X\to X) = G_{\mu \nu}(X)~.
\eea
Notice that the representations in (\ref{D-rep}) is unique up to a
single Lorentz transformation which may be taken to be the one
relating the objects, say, at $X$ in the two expressions. But
according to the definition of covariant derivative
in (\ref{cov-der-vier}) a Lorentz transformation matrix is invariant
under parallel transport and therefore the same transformation relates
the objects at $\tilde X$. However, when $D_{\mu \nu}(X, \tilde X)$ acts on a
vector this ambiguity gets canceled by the same in defining the
Lorentz component of that vector.

We now generalize the above discussion to a string. We consider two
nearby string-embeddings $X(\s)$ and $\tilde X(\s)$ in
physical spacetime, both parametrized by $\sigma$, such that the pair
of points on the two strings at the same value of $\s$ are connected by
a unique geodesic. This leads to the definition of the geodetic parallel displacement bi-vector $D_{\mu \nu}(X(\s), \tilde X(\s))$ for the string.
The infinite-dimensional analogue of the same is given by the general map
(\ref{th-rule}),
\bea
\Del_{ij}(x,\tilde x) &=& \oint {d\s \over 2\pi}~ D_{\mu \nu}(X(\s), \tilde X(\s))
e^{i(m+n)\s} ~, \cr
&=& e_{~i}^{\hat i}(\tilde x \to x)\eta_{\hat i \hat j}e^{\hat j}_{~j}(\tilde x)
~, \cr
&=& e_{~i}^{\hat i}(x)\eta_{\hat i \hat j}e^{\hat j}_{~j}(x\to \tilde x) ~,
\label{Del-def}
\eea
where the infinite-dimensional analogues of the vielbeins and
Minkowski metric are given by,
\bea
e^{\hat i}_{~i}(x) &=& \oint {d\s \over 2\pi}~ E^{\hat
  \mu}_{~\mu}(X(\s)) e^{i(m-\hat m)\s}~ , \cr
\eta_{\hat i \hat j} &=& \oint {d\s\over 2\pi}~ \eta_{\hat \mu \hat \nu}
e^{i(m+n)\s}= \eta_{\hat \mu \hat \nu} \dt_{m+n,0}~,
\label{vielbein-inf}
\eea
respectively. This way all the relevant equations that we discussed in the context
of physical spacetime have infinite-dimensional analogues. For
example,  the analogue of eq.(\ref{cov-der-vier}), which we use extensively in our computations, is given by,
\bea
\nabla_i e^{\hat i}_{~j} = \del_i e^{\hat i}_{~j} -
\gamma^k_{ij}e^{\hat i}_{~k} = -\omega_{i ~~\hat j}^{~\hat i} e^{\hat j}_{~j}~,
\label{cov-der-vier-inf}
\eea
where,
\bea
w_{i \hat i \hat j} = \oint {d\s \over 2\pi}~\Omega_{\mu \hat \mu
  \hat \nu}(X(\s)) e^{i(m+\hat m +\hat n)}~.
\label{omega}
\eea

\section{Tensor representation of $\hat \pi_k$}
\label{a:pi-rep}

Here we will derive the tensor representation of $\pih_k$ as given in (\ref{pi-tensor-rep}). We follow the basic line of argument in \cite{dewitt52}. In particular, we demand that the following operator equations hold as matrix elements,
\bea
[\hat x^l, \pih_k] = i \dt^l_k~, &&  [\pih_k,\pih_l] = 0~,
\label{pi-comm}
\eea
\bea
\pih^{\dagger}_k &=& \pih^{\star}_{\bar k}~.
\label{pi-hermiticity}
\eea
Using the first commutator in (\ref{pi-comm}) one gets the following condition,
\bea
(x-\tilde x)^l \la{}_{i_1\cdots i_n};x|\pih_k|{}_{j_1\cdots j_{\tilde n}};\tilde x\ra
&=& i \dt^l_k \dt_{n,\tilde n} \Del_{i_1j_1}(x,\tilde x) \cdots \Del_{i_nj_n}(x,\tilde x) \dt(x,\tilde x)~. \cr &&
\eea
Using the identities,
\bea
(x-\tilde x)^l \del_k \dt(x,\tilde x) &=& -\dt^l_k \dt(x,\tilde x)~, \cr
\del_k \dt(x,\tilde x) &=& - (\tilde \del_k + \gamma_k(x)) \dt(x,\tilde x)~,
\label{del-identities}
\eea
we write down the general solution in the form as given in (\ref{pi-tensor-rep-ele}) with $\Del_{i_1\cdots i_n j_1\cdots j_n}(x,\tilde x)$ as given in (\ref{ortho}). The hermiticity condition in (\ref{pi-hermiticity}) is
easily verified by using the expression of $F_{i_1\cdots i_n j_1\cdots j_n k}(x,\tilde x)$ as given in
(\ref{F-n-def}).
\bea
\la {}_{i_1\cdots i_n};x|\pih_k^{\dagger}|{}_{j_1\cdots j_{\tilde n}};\tilde x\ra &=&
\la {}_{\bar j_1\cdots \bar j_{\tilde n}};\tilde x|\pih_k|{}_{\bar i_1\cdots \bar
  i_n};x \ra^* ~,\cr
&=& i \dt_{n\tilde n}\lt[\Del_{j_1\cdots j_n i_1\cdots i_n}(\tilde x,x) \tilde \del_{\bar
  k} \dt(x,\tilde x) + F^*_{{\bar j}_1\cdots {\bar j}_n {\bar i}_1\cdots {\bar i}_n 
k}(\tilde x,x)
\dt(x,\tilde x) \rt] ~, \cr
&=&-i \dt_{n\tilde n}\lt[\Del_{i_1\cdots i_n j_1 \cdots j_n}(x, \tilde x) (\del_{\bar k} + \gamma_{\bar k}(x) )\dt(x, \tilde x) \rt. \cr
&& \lt. + F_{i_1\cdots i_n j_1 \cdots j_n \bar k}(x,\tilde x) \dt(x,\tilde x) \rt]~,  \cr 
&=& \la{}_{i_1\cdots i_n};x|\pih^{\star}_{\bar k}|{}_{j_1\cdots j_n};\tilde x\ra ~,
\eea
where in the second step we have used the reality property (\ref{reality-Del}). In the third step we have used,
\bea
\Del_{i_1\cdots i_n j_1\cdots j_n}(x,\tilde x) &=& \Del_{j_1\cdots
  j_n i_1\cdots i_n}(\tilde x,x)~,
\label{symm-Del}
\eea 
the second equation in (\ref{del-identities}) and,
\bea
F^*_{i_1\cdots i_n j_1 \cdots j_n k}(x,\tilde x) &=& F_{{\bar i}_1\cdots {\bar i}_n {\bar j}_1 \cdots {\bar j}_n \bar k}(x,\tilde x) ~, \cr
F_{i_1\cdots i_n j_1\cdots j_n k}(x,\tilde x) \dt(x,\tilde x) &=& - F_{j_1\cdots j_n i_1\cdots i_n k}(\tilde x, x) \dt(x, \tilde x)~.
\label{reality-symm-F}
\eea

We will verify the second commutator in (\ref{pi-comm}) between two arbitrary invariant tensor states using the expressions in (\ref{pi-tensor-rep}) instead of the position eigenstates. This will make the expressions look more transparent. To this end we
first show how one gets the results in (\ref{pi-tensor-rep}) from
(\ref{pi-tensor-rep-ele}). We first expand $\Del_{ij}(x,\tilde x)$ about $\tilde x$,
\bea
\Del_{ij}(x,\tilde x) &=& g_{ij}(\tilde x) +(x-\tilde x)^k \gamma^{l}_{ki}(\tilde x)g_{lj}(\tilde x)+ \cdots~.
\eea
This is obtained by expanding the vielbein $e^{\hat i}_{~i}(\tilde x \to x)$ in the second equation of (\ref{Del-def}) up to first order. Using this one finds,
\bea
 \Del_{i_1\cdots i_n j_1\cdots j_n}(x,\tilde x) \del_k \dt(x,\tilde x) &=&
 g_{i_1\cdots i_n j_1\cdots j_n}(\tilde x) \del_k \dt(x,\tilde x) -
 \gamma_{i_1\cdots i_n j_1\cdots j_n k}(x) \dt(x,\tilde x) ~,  \cr &&
\label{Del-del-dt}
\eea
where $g_{i_1\cdots i_n j_1\cdots j_n}$ and $\gamma_{i_1\cdots i_n
  j_1\cdots j_n k}$ are as given in (\ref{g-n-def}) and
(\ref{gamma-n-def}) respectively.  Interchanging $x \leftrightarrow
\tilde x$ and $i\leftrightarrow j$ in (\ref{Del-del-dt}) and using (\ref{symm-Del}) and,
\bea
g_{i_1\cdots i_n j_1\cdots j_n}(x) &=& g_{j_1\cdots j_n i_1\cdots i_n}(x)~,
\eea
one establishes,
\bea
\Del_{i_1\cdots i_n j_1\cdots j_n}(x,\tilde x) \tilde \del_k \dt(x,\tilde x) &=&
 g_{i_1\cdots i_n j_1\cdots j_n}(x) \tilde \del_k \dt(x,\tilde x)
 -\gamma_{j_1\cdots j_n i_1\cdots i_n k}(x) \dt(x,\tilde x) ~. \cr &&
\label{Del-del'-dt}
\eea
Finally it is straightforward to show,
\bea
F_{i_1\cdots i_n j_1\cdots j_n k}(x,\tilde x) \dt(x, \tilde x) &=& \omega_{i_1\cdots i_n j_1\cdots j_n k}(x) \dt(x, \tilde x)~.
\label{F-n-expand}
\eea
Substituting the results
(\ref{Del-del-dt}), (\ref{Del-del'-dt}) and (\ref{F-n-expand}) in
(\ref{pi-tensor-rep-ele}) one finds the results (\ref{pi-tensor-rep}).

Next we will consider the matrix element of the second commutator in
(\ref{pi-comm}) between two arbitrary tensor states $|\chi_{(n)}\ra$ and
$|\psi_{(n)}\ra$ of rank $n$. In order to do such calculations we need the following results:
\bea
\la \chi_{(n)}|\hat \pi_k |{}_{j_1\cdots j_n}; x \ra &=& i
\lt[(\nabla_k +\gamma_k(x)) \bar \chi_{j_1\cdots j_n}(x)
+\bar \chi^{i_1\cdots i_n}(x) \omega_{i_1\cdots i_n j_1\cdots j_n k}(x)
\rt] ~,
%\label{chi-pih-tensorx} \\
\cr
\la{}_{i_1\cdots i_n};x|\hat \pi_k|\psi_{(n)} \ra &=& -i
\lt[\nabla_k \psi_{i_1\cdots i_n}(x) -\omega_{i_1\cdots i_n j_1\cdots
  j_n k}(x) \psi^{j_1\cdots j_n}(x) \rt]~.
\label{tensorx-pih-psi}
\eea
Such results are derived using the expressions in
(\ref{pi-tensor-rep}) and the following relations,
\bea
\del_k g_{i_1\cdots i_n j_1\cdots j_n} &=& \gamma_{i_1\cdots
  i_n j_1\cdots j_n k} + \gamma_{j_1\cdots j_n i_1\cdots i_n k}~, \cr
\nabla_k \psi_{i_1\cdots i_n} &=& \del_k \psi_{i_1\cdots i_n} - \gamma_{i_1\cdots i_n j_1\cdots j_n k}\psi^{j_1\cdots j_n}~. 
\eea
We will now calculate $\la\chi |\hat \pi_k \hat \pi_l|\psi \ra$ by first inserting the completeness relation (\ref{compl}) between the two $\pih$'s and then using the results (\ref{tensorx-pih-psi}). To
get to the final expression we need to go through certain manipulations that are generally involved in the computations of the matrix elements carried out in this paper. We give the details of the present calculation in a series of steps below,
\bea
\la\chi_{(n)} |\hat \pi_k \hat \pi_l|\psi_{(n)} \ra
&=& \int dw~ \la\chi_{(n)} |\hat \pi_k |{}_{k_1\cdots k_n};x\ra
g^{k_1 l_1}(x) \cdots g^{k_n l_n}(x) \la {}_{l_1\cdots l_n};x|\hat
\pi_l|\psi_{(n)}\ra ~, \cr
&=& \int dw~ \lt\{(\nabla_k+\gamma_k)\bar \chi_{k_1\cdots
  k_n}+ \bar \chi^{i_1\cdots i_n}\omega_{i_1\cdots i_n k_1\cdots k_n k}
\rt\} \cr
&& \lt(\nabla_l\psi^{k_1\cdots k_n} -\omega^{k_1\cdots k_n}_{~~~~~~l_1\cdots l_n l}  \psi^{l_1\cdots l_n}\rt) ~, \cr
&=& \int dw~ \lt\{(\del_k +\gamma_k) \bar \chi_{k_1\cdots k_n} -(\gamma^{k'}_{kk_1} \bar \chi_{k'k_2\cdots k_n} +
\gamma^{k'}_{kk_2} \bar \chi_{k_1k'k_3\cdots k_n} + \cdots) \rt. \cr
&& \lt. + \bar \chi^{i_1\cdots i_n}\omega_{i_1\cdots i_n k_1\cdots k_n k}\rt\} \lt(\nabla_l\psi^{k_1\cdots k_n} -\omega^{k_1\cdots k_n}_{~~~~~~l_1\cdots l_n l} \psi^{l_1\cdots l_n}\rt) ~, \cr
&=& - \int dw~ \lt[\bar \chi_{k_1\cdots k_n} \del_k\lt(\nabla_l
\psi^{k_1\cdots k_n}-\omega^{k_1\cdots k_n}_{~~~~~~l_1\cdots l_n l}
\psi^{l_1\cdots l_n} \rt) \rt. \cr
&& \lt. + \bar \chi_{k_1\cdots k_n}\lt\{\gamma^{k_1}_{kk'}\lt(\nabla_l
\psi^{k'k_2\cdots k_n}-\omega^{k'k_2\cdots k_n}_{~~~~~~~~l_1\cdots l_n l}
\psi^{l_1\cdots l_n} \rt) \rt. \rt. \cr
&& \lt. + \gamma^{k_2}_{kk'}\lt(\nabla_l
\psi^{k_1k'k_2\cdots k_n}-\omega^{k_1k'k_2\cdots k_n}_{~~~~~~~~~~l_1\cdots l_n l}
\psi^{l_1\cdots l_n} \rt)  \cdots \rt\}  \cr
&& \lt. -\bar \chi^{i_1\cdots i_n} \omega_{i_1\cdots i_n k_1\cdots k_n k}
\lt(\nabla_l \psi^{k_1\cdots k_n} - \omega^{k_1\cdots
  k_n}_{~~~~~~l_1\cdots l_n l} \psi^{l_1\cdots l_n} \rt)\rt]  ~, \cr
&=& - \int dw~ \lt[\bar \chi_{k_1\cdots k_n}
\nabla_k\lt(\nabla_l \psi^{k_1\cdots k_n}-\omega^{k_1\cdots k_n}_{~~~~~~l_1\cdots l_n l} \psi^{l_1\cdots l_n} \rt) \rt. \cr
&& \lt. + \bar \chi_{k_1\cdots k_n} \gamma^{l'}_{kl}\lt(\nabla_{l'}
\psi^{k_1\cdots k_n}-\omega^{k_1\cdots k_n}_{~~~~~~l_1\cdots l_n l'}
\psi^{l_1\cdots l_n} \rt) \rt. \cr
&& \lt. - \bar \chi^{i_1\cdots i_n} \omega_{i_1\cdots i_n k_1\cdots k_n k}\lt(\nabla_l \psi^{k_1\cdots k_n}-\omega^{k_1\cdots k_n}_{~~~~~~l_1\cdots l_n l}
\psi^{l_1\cdots l_n} \rt)\rt] ~, \cr
&=& - \int dw~ \lt[\bar \chi_{k_1\cdots k_n} \nabla_k \nabla_l
\psi^{k_1\cdots k_n} - \bar \chi_{k_1\cdots k_n} \nabla_k
\omega^{k_1\cdots k_n}_{~~~~~~l_1\cdots l_n l} \psi^{l_1\cdots l_n}
\rt. \cr
&& \lt. + \bar \chi^{i_1\cdots i_n} \omega_{i_1\cdots i_n k_1\cdots k_n k} \omega^{k_1\cdots k_n}_{~~~~~~l_1\cdots l_n l} \psi^{l_1\cdots l_n} \rt. \cr
&& \lt. + \bar \chi_{k_1\cdots k_n} \gamma^{l'}_{kl}\lt(\nabla_{l'} \psi^{k_1\cdots k_n}-\omega^{k_1\cdots k_n}_{~~~~~~l_1\cdots l_n l'} \psi^{l_1\cdots l_n} \rt) \rt. \cr
&& \lt. - \bar \chi_{k_1\cdots k_n} \lt(\omega^{k_1\cdots k_n}_{~~~~~~l_1\cdots l_nl} \nabla_k + k\leftrightarrow l \rt) \psi^{j_1\cdots j_n}\rt]~.
\label{manipulation}
\eea
The last two terms in the above expression are symmetric in $k$ and
$l$. Therefore they do not contribute to the commutator which is given by,
\bea
\la\chi_{(n)}|[\hat \pi_k, \hat \pi_l] |\psi_{(n)} \ra = - \int dw~ V~,
\eea
where,
\bea
V &=& \bar \chi_{k_1\cdots k_n} [\nabla_k, \nabla_l]\psi^{k_1\cdots k_n}
- \bar \chi_{k_1\cdots k_n}\lt(\nabla_k \omega^{k_1\cdots
  k_n}_{~~~~~~l_1\cdots l_nl} - k\leftrightarrow l \rt)
\psi^{l_1\cdots l_n} \cr
&&  + \bar \chi^{i_1\cdots i_n}\lt(\omega_{i_1\cdots i_n k_1\cdots
  k_n k}\omega^{k_1\cdots k_n}_{~~~~~~j_1\cdots j_n l} - k \leftrightarrow l \rt)
\psi^{j_1\cdots j_n} ~.
\label{V-def}
\eea
Below we will show that $V$ vanishes. Using,
\bea
[\nabla_k, \nabla_l]\psi^{k_1\cdots k_n} =
r^{k_1}_{~k'kl}\psi^{k'k_2\cdots k_n} +
r^{k_2}_{~k'kl}\psi^{k_1k'k_3\cdots k_n} + \cdots ~,
\eea
$r^i_{~jkl}$ being the Riemann tensor, one may write the first term in (\ref{V-def}) as,
\bea
\bar \chi_{k_1\cdots k_n}[\nabla_k, \nabla_l] \psi^{k_1\cdots k_n} =
\bar \chi^{k_1\cdots k_n} r_{k_1\cdots k_n l_1\cdots l_n kl}\psi^{l_1\cdots l_n}~,
\label{chi-r-psi}
\eea
where,
\bea
r_{k_1\cdots k_n l_1\cdots l_n kl} = r_{k_1l_1kl}g_{k_2l_2}\cdots
g_{k_nl_n} + g_{k_1l_1}r_{k_2l_2kl}g_{k_3l_3}\cdots g_{k_nl_n} +
\cdots ~.
\eea
To calculate the second term in (\ref{V-def}) one first derives, following
the standard argument \cite{wald84},
\bea
\nabla_k w_{lij} - k\leftrightarrow l &=& r_{ijkl} + d_{ijkl}~,  \cr
d_{ijkl} &=& w_{kii'}w^{~i'}_{l~~j} - k\leftrightarrow l ~,
\eea
using which one may write,
\bea
\nabla_k \omega_{k_1\cdots k_n l_1\cdots l_n l} - k \leftrightarrow l
&=& r_{k_1\cdots k_n l_1\cdots l_n kl} + d_{k_1\cdots k_n l_1\cdots l_n kl}
\label{dalpha}
\eea
where we have defined,
\bea
d_{i_1\cdots i_n,j_1\cdots j_n,kl} &\equiv & d_{i_1j_1kl} g_{i_2j_2}
\cdots g_{i_nj_n} + g_{i_1j_1} d_{i_2j_2kl} g_{i_3j_3} \cdots
g_{i_nj_n} + \cdots
\eea
We note down the following anti-symmetry properties for future use,
\bea
r_{i_1\cdots i_n j_1\cdots j_n kl} &=& -r_{j_1\cdots j_n i_1\cdots i_n
  kl} = - r_{i_1\cdots i_n j_1\cdots j_n lk} ~, \cr
d_{i_1\cdots i_n j_1\cdots j_n kl} &=& -d_{j_1\cdots j_n i_1\cdots i_n
  kl} = - d_{i_1\cdots i_n j_1\cdots j_n lk}~.
\label{r-d-anti-symmetry}
\eea
The last term in (\ref{V-def}) is calculated in the following way,
\bea
\omega_{i_1\cdots i_n k_1\cdots k_n k} \omega^{k_1\cdots
  k_n}_{~~~~~~j_1\cdots j_n l} - k\leftrightarrow l
&=& \lt(\omega_{ki_1k_1} g_{i_2k_2} \cdots g_{i_nk_n}  + g_{i_1k_1}
\omega_{ki_2k_2} g_{i_3k_3} \cdots g_{i_nk_n} + \cdots \rt) \cr &&
\lt(\omega^{~k_1}_{l~~j_1} g^{k_2}_{~j_2} \cdots g^{k_n}_{~j_n} +
g^{k_1}_{~j_1} \omega^{~k_2}_{l~~j_2} g^{k_3}_{~j_3} \cdots
g^{k_n}_{~j_n} + \cdots \rt) - k\leftrightarrow l~, \cr
&=& {\cal D}_{i_1\cdots i_n,j_1\cdots j_n,kl} + {\cal C}_{i_1\cdots
  i_n,j_1\cdots j_n,kl} - k\leftrightarrow l ~,
\label{alpha-square-pre}
\eea
where the {\it diagonal} and the {\it cross} terms are,
\bea
{\cal D}_{i_1\cdots i_n,j_1\cdots j_n,kl} &=&
\lt(\omega_{ki_1k_1}\omega^{~k_1}_{l~~j_1}\rt) g_{i_2j_2} \cdots
g_{i_nj_n} \cr
&& + g_{i_1j_1} \lt(\omega_{ki_2k_2}\omega^{~k_2}_{l~~j_2}\rt)
g_{i_2j_2} \cdots g_{i_nj_n} + \cdots ~, \cr
{\cal C}_{i_1\cdots i_n,j_1\cdots j_n,kl} &=&
\lt\{(\omega_{ki_1j_1})(\omega_{li_2j_2}) g_{i_3j_3} \cdots g_{i_nj_n}
+ k\leftrightarrow l \rt\} \cr
&& + \lt\{(\omega_{ki_1j_1}) g_{i_2j_2} (\omega_{li_3j_3}) g_{i_4j_4}
\cdots g_{i_nj_n} + k\leftrightarrow l \rt\} + \cdots ~,
\eea
respectively. The cross term, being symmetric under $k \leftrightarrow
l$, does not contribute to (\ref{alpha-square-pre}). Therefore,
\bea
\omega_{i_1\cdots i_n k_1\cdots k_n k} \omega^{k_1\cdots
  k_n}_{~~~~~~j_1\cdots j_n l} - k\leftrightarrow l = d_{i_1\cdots i_n
  j_1\cdots j_n kl}~.
\label{alpha-square}
\eea

Using the results in (\ref{chi-r-psi}), (\ref{dalpha}) and (\ref{alpha-square}) in (\ref{V-def}) one shows that $V$ vanishes implying,
\bea
\la\chi_{(n)}|[\hat \pi_k,\hat \pi_l]|\psi_{(n)}\ra = 0~.
\eea
The above calculations show why the $F$-term in
eqs.(\ref{pi-tensor-rep-ele}) or  the spin connection term in
eqs.(\ref{pi-tensor-rep}) is essential. This is needed to cancel the
Riemann tensor contributions which will otherwise survive in the
calculation of the above commutator.

\section{GCT as a unitary transformation}
\label{a:unitary}

It should be possible to understand GCT, as introduced below eq.(\ref{delta}), as a usual unitary transformation in quantum mechanics. This has been explained for spin-zero representation in \cite{dewitt57}. Here we will generalize the argument to incorporate the tensor representations. 

The rank-$n$ tensor wavefunction and the position eigenstate in eq.(\ref{psi-n}) transform under GCT in the following way,
\bea
\psi^{i_1\cdots i_n}(x) &\to& \psi'^{i_1\cdots i_n}(x')
= \lambda^{i_1}_{~j_1}(x) \cdots \lambda^{i_n}_{~j_n}(x)
\psi^{j_1\cdots j_n}(x)~, \cr
|{}_{i_1i_2\cdots i_n} ; x\ra &\to& |{}_{i_1i_2\cdots i_n} ; x'\ra' =
\lambda_{i_1}^{~j_1}(x) \lambda_{i_2}^{~j_2}(x) \cdots
\lambda_{i_n}^{~j_n}(x) |{}_{j_1 j_2\cdots i_n} ; x\ra ~.
\label{GCT-psi-ket}
\eea
Using (\ref{GCT}) and (\ref{GCT-psi-ket}) one finds, as expected,
\bea
\hat x'^i |{}_{i_1\cdots i_n}; x'\ra' &=& x'^i |{}_{i_1\cdots i_n}; x'\ra'~.
\label{f-eigenstate}
\eea
Let us now consider the following unitary operator,
\bea
\hat U &=& e^{{i\over 2} \hat V}~, \cr
\hat V &=& \pih^{\star}_i v^i(\xh)+ v^i(\xh) \pih_i~,
\label{U-def}
\eea
where the $\hat \pi$-operators are defined in (\ref{pih-def}) and  $v^i(x)$ is a vector field such that,
\bea
v^i(x)^* = v^{\bar i}(x)~.
\eea
The above definition implies that $\hat V$ is both hermitian and GCT invariant. The latter can be checked easily by using the transformation property in (\ref{GCT-pi}).
At the infinitesimal level $\hat U$ produces the following coordinate transformation,
\bea
\hat x'^i(\xh) &=& \hat U \xh^i \hat U^{\dagger} = \xh^i + v^i(\xh)~.
\label{fi-U}
\eea
The question that one would like to ask is how the state $\hat U|{}_{i_1\cdots i_n}; x\ra$ is related to $|{}_{i_1\cdots i_n}; x\ra'$. Notice that both these states have the same eigenvalue for $\hat x'^i$, namely $x^i$.

To answer the above question we first write the infinitesimal version of  (\ref{GCT-psi-ket}) in the following way,
\bea
|{}_{i_1\cdots i_n}; x-v(x)\ra &=& (\dt_{i_1}^{j_1}\dt_{i_2}^{j_2}\cdots \dt_{i_n}^{j_n} + \nabla_{i_1}v^{j_1} \dt_{i_2}^{j_2}\cdots \dt_{i_n}^{j_n} + \dt_{i_1}^{j_1} \nabla_{i_2} v^{j_2} \dt_{i_3}^{j_3} \cdots \dt_{i_n}^{j_n} + \cdots  ) |{}_{j_1\cdots j_n}; x\ra'~, \cr &&
\eea
where the state appearing on the left hand side has a covariant Taylor expansion in the following sense,
\bea
\la {}_{i_1\cdots i_n};x-v(x)|\psi_{(n)} \ra = \psi_{i_1\cdots i_n}(x) - v^k(x) \nabla_k \psi_{i_1\cdots i_n}(x)~,
\eea
where $|\psi_{(n)}\ra$ is an arbitrary rank-$n$ tensor state as in (\ref{psi-n}).
Below we will show that the following relation is true,
\bea
\hat U |{}_{i_1\cdots i_n};x\ra &=& N_{i_1\cdots i_n}^{~~~~~j_1\cdots j_n}(x) |{}_{j_1\cdots j_n}; x- v(x)\ra ~, 
\label{Ustate-state}
\eea
where the normalization is given by,
\bea
N_{i_1\cdots i_n j_1\cdots j_n} &=& g_{i_1\cdots i_n j_1\cdots j_n} + \dt N_{i_1\cdots i_n j_1\cdots j_n} ~, \cr
\dt N_{i_1\cdots i_n j_1\cdots j_n} &=& -{1\over 2} \nabla_k v^k g_{i_1\cdots i_n j_1\cdots j_n} + \omega_{i_1\cdots i_nj_1 \cdots j_n k} v^k~.
\label{norm}
\eea

To prove (\ref{Ustate-state}, \ref{norm}) we will first calculate $\la {}_{i_1\cdots i_n};x|\hat U^{\dagger}|\psi_{(n)}\ra$ independently by using (\ref{Ustate-state}) and 
(\ref{U-def}) and then compare the results. Using (\ref{Ustate-state}) one gets,
\bea
\la{}_{i_1\cdots i_n};x|\hat U^{\dagger}|\psi_{(n)}\ra &=& N_{\bar i_1 \cdots \bar i_n}^{*~~~~\bar j_1\cdots \bar j_n}(x)(\psi_{j_1\cdots j_n}(x) - v^k(x)\nabla_k \psi_{j_1\cdots j_n}(x))~, \cr 
&=& [1- v^k(x) \nabla_k] \psi_{i_1\cdots i_n}(x) + \dt N_{\bar i_1 \cdots \bar i_n}^{*~~~~\bar j_1 \cdots \bar j_n}\psi_{j_1\cdots j_n}(x)~,
\label{state-U-psi1}
\eea
where in the second step we have used the first equation in (\ref{norm}). On the other hand using (\ref{U-def}) one finds,
\bea
\la {}_{i_1\cdots i_n};x|\hat U^{\dagger}|\psi_{(n)}\ra &=& [1-{1\over 2}\nabla_k v^k(x)- v^k(x)\nabla_k ]\psi_{i_1\cdots i_n}(x) + \omega_{i_1\cdots i_n j_1\cdots j_n k}(x)v^k(x) \psi^{j_1\cdots j_n}(x) ~, \cr &&
\label{state-U-psi2}
\eea
where we have used the representation (\ref{pi-tensor-rep}). Finally, comparing (\ref{state-U-psi1}) and (\ref{state-U-psi2}) we arrive at the second equation in (\ref{norm}).

\section{Derivation of DWV algebra}
\label{a:alg}

Here we will reproduce the results in (\ref{DWValg}, \ref{DWVanomaly}) in tensor representation. This will be done by showing that the results hold
as matrix elements between two arbitrary tensor states\footnote{To
  simplify our notations we remove the subscript $(n)$ from the states
in this appendix.} $|\chi \ra$ and $|\psi\ra$ of rank $n$. We begin by
quoting certain results that will be needed as we go along.
\bea
\la\chi|\hat K_{(i)}|{}_{i_1\cdots i_n};x\ra
&=& (-i)^2 \lt[\nabla_{(i)}^2 \bar \chi_{i_1\cdots i_n} +
2\nabla^{k+i}\bar \chi^{j_1\cdots j_n}\omega_{j_1\cdots j_n i_1\cdots i_n
  k} + \bar \chi^{j_1\cdots j_n}\nabla^{k+i}\omega_{j_1\cdots j_n
  i_1\cdots i_n k} \rt. \cr
&& \lt. + \bar \chi^{j_1\cdots j_n}\omega_{j_1\cdots j_n k_1\cdots k_n
  k}\omega^{k_1\cdots k_n~~~~~k+i}_{~~~~~~i_1\cdots i_n} \rt]~,
\label{chi-K-tensor} \\
\la{}_{i_1\cdots i_n};x|\hat K_{(i)}|\psi\ra
&=& (-i)^2 \lt[\nabla_{(i)}^2 \psi_{i_1\cdots i_n} -
\nabla^{k+i}\omega_{i_1\cdots i_n j_1\cdots j_n k}\psi^{j_1\cdots j_n}
-2\omega_{i_1\cdots i_n j_1\cdots j_n k} \nabla^{k+i}\psi^{j_1\cdots j_n} \rt. \cr
&& \lt. + \omega_{i_1\cdots i_n k_1\cdots k_n k}\omega^{k_1\cdots
  k_n~~~~~k+i}_{~~~~~~j_1\cdots j_n} \psi^{j_1\cdots j_n}\rt] ~,
\label{tensor-K-psi} \\
\la\chi|\hat Z^L_{(i)}|{}_{i_1\cdots i_n};x\ra &=& -i
\lt(-\nabla_k \bar \chi_{i_1\cdots i_n}- \bar \chi^{j_1\cdots
  j_n}\omega_{j_1\cdots j_n i_1\cdots i_n k} \rt) a^{k+i}~,
\label{chi-ZL-tensor} \\
\la{}_{i_1\cdots i_n};x|\hat Z^L_{(i)}|\psi\ra &=& -i
\lt[\nabla_k(a^{k+i}\psi_{i_1\cdots i_n}) - \omega_{i_1\cdots i_n
  j_1\cdots j_n k} a^{k+i} \psi^{j_1\cdots j_n} \rt] ~.
\label{tensor-ZL-psi} \\
\la\chi|\hat Z^R_{(i)}|{}_{i_1\cdots i_n};x\ra &=& -i
\lt[-\nabla_k(\bar \chi_{i_1\cdots i_n} a^{k+i}) - \bar \chi^{j_1\cdots
  j_n}\omega_{j_1\cdots j_n i_1\cdots i_n k}a^{k+i} \rt] ~,
\label{chi-ZR-tensor} \\
\la{}_{i_1\cdots i_n};x|\hat Z^R_{(i)}|\psi\ra &=& -i
a^{k+i}\lt[\nabla_k \psi_{i_1\cdots i_n} - \omega_{i_1\cdots i_n
  j_1\cdots j_n k}\psi^{j_1\cdots j_n} \rt] ~.
\label{tensor-ZR-psi}
\eea
To proceed with the computation of DWV algebra we first introduce
certain notations. We break each of the relevant commutators into several
terms in the following way,
\bea
_{\chi}\la [\hat L_{(i)}, \hat L_{(j)}]\ra_{\psi} &=& {1\over 16} [ T^{K_{(i)}K_{(j)}} +
T^{Z_{(i)}Z_{(j)}} - (T^{K_{(i)}Z_{(j)}} - i \leftrightarrow j ) +
(T^{K_{(i)}V_{(j)}} - i \leftrightarrow j) \cr
&& - (T^{Z_{(i)}V_{(j)}} - i \leftrightarrow j) ]~, \cr
_{\chi}\la [\hat{ \tilde L}_{(i)}, \hat{ \tilde L}_{(j)}]\ra_{\psi}
&=& {1\over 16}
[T^{K_{(\bar i)}K_{(\bar j)}} + T^{Z_{(\bar i)} Z_{(\bar j)}} +
(T^{K_{(\bar i)}Z_{(\bar j)}}  -
\bar i \leftrightarrow \bar j) + (T^{K_{(\bar i)}V_{(\bar j)}} - \bar i \leftrightarrow \bar j)
\cr
&& + (\hat T^{Z_{(\bar i)}V_{(\bar j)}} - \bar i \leftrightarrow \bar j ) ]~, \cr
_{\chi}\la[\hat L_{(i)}, \hat{\tilde L}_{(j)}]\ra_{\psi} &=&{1\over
  16}[ T^{K_{(i)}K_{(\bar j)}} - T^{Z_{(i)}Z_{(\bar j)}} +
(T^{K_{(i)}Z_{(\bar j)}} + i \leftrightarrow \bar j) +
(T^{K_{(i)}V_{(\bar j)}} - i \leftrightarrow \bar j) \cr
&& - (T^{Z_{(i)}V_{(\bar j)}} + i \leftrightarrow \bar j )]~, \cr &&
\label{DWValg-parts}
\eea
where $_{\chi}\la \cdots \ra_{\psi} = \la \chi |\cdots |\psi \ra$ and,
\bea
T^{AB} = {}_{\chi}\la [\hat A, \hat B]\ra_{\psi}~.
\label{TAB-def}
\eea
Terms involving the operator $\hat Z_{(i)}$ are further divided into
components in the following way,
\bea
T^{Z_{(i)}Z_{(j)}} &=&  T^{Z^L_{(i)}Z^L_{(j)}} +
T^{Z^R_{(i)}Z^R_{(j)}} + (T^{Z^L_{(i)}Z^R_{(j)}} - i\leftrightarrow
j)~,
\label{Z-Z-comm} \\
T^{K_{(i)}Z_{(j)}} &=& T^{K_{(i)}Z^L_{(j)}} + T^{K_{(i)}Z^R_{(j)}}~,
\label{K-Z-comm}\\
T^{Z_{(i)}V_{(j)}} &=& T^{Z^L_{(i)}V_{(j)}} + T^{Z^R_{(j)}V_{(j)}}~.
\label{Z-V-comm}
\eea
Each of the commutator terms receives contributions at different
orders in $\omega_{i_1\cdots i_nj_1\cdots j_nk}$,
\bea
T^{AB}&=& (-i)^r \lt[\bar O _{\chi \psi}^{AB} + \bar L _{\chi
  \psi}^{AB} + \bar Q _{\chi \psi}^{AB} + \bar C _{\chi
  \psi}^{AB}\rt]~,
\label{TAB-parts}
\eea
where $r$ is the number of $\pih$ operator involved in the computation
and $\bar O$, $\bar L$, $\bar Q$ and $\bar C$ denote the terms which
are zeroth, first, second and third order in $\omega$
respectively. We have made ordering of $\chi$ and $\psi$ explicit in the notations of such contributions as we will exploit certain symmetry under $\bar \chi
\leftrightarrow \psi$ to compute a particular set of terms. Although
for other terms it is not needed, we use a uniform notation. Similar
contributions for a matrix element of product of two operators will
be denoted by the same symbols, but without the bar\footnote{The quantity ${}_{\chi}\la \hat K_{(i)} \hat K_{(j)} \ra_{\psi}$ receives contribution at
the quartic order, but we will not have to calculate it explicitly.},
\bea
{}_{\chi}\la \hat A \hat B \ra_{\psi} &=& (-i)^r \lt[O^{AB}_{\chi \psi} +
L^{AB}_{\chi \psi} + Q^{AB}_{\chi \psi} + C^{AB}_{\chi \psi} \rt]~.
\label{AB-matrix-parts}
\eea
Below we will compute each of the terms appearing in
(\ref{DWValg-parts}, \ref{Z-Z-comm}, \ref{K-Z-comm}, \ref{Z-V-comm}) separately.

\vspace{.1in}
\noindent
\underline{\large $T^{K_{(i)}K_{(j)}}$} \\
Using (\ref{chi-K-tensor}) and (\ref{tensor-K-psi}) one can explicitly
compute all the terms in ${}_{\chi}\la \hat K_{(i)} \hat K_{(j)}\ra_{\psi}$. 
In \cite{pm0912} it was shown that in spin-zero representation such terms are symmetric under $i\leftrightarrow j$ by using the shift property in
(\ref{shift}). In the present case the wavefunctions themselves carry tensor indices and a direct application of such an argument would require the wavefunctions also to possess similar shift property. According to our general definition of tensor wavefunctions which has been explained below (\ref{scope}), these objects may in general involve derivatives of $a^i(x)$. Since the general form of $u^{i_1 j_1\cdots}_{i_2 j_2\cdots }$ (see discussion above (\ref{replace})), for which the shift property in (\ref{shift}) holds true, does not allow such factors, we do not want to assume the shit property for the wavefunctions. However, as will be discussed below, a slightly modified argument may be used to show that all the terms in ${}_{\chi}\la \hat K_{(i)} \hat K_{(j)}\ra_{\psi}$ are indeed symmetric in $i$ $j$ without assuming any special property for the wavefunctions. This would therefore imply,
\bea
T^{K_{(i)}K_{(j)}} &=& 0~.
\label{T-KK}
\eea

We begin with the ${\cal O}(\omega^0)$-term. By using integration by parts we write it in the following form,
\bea
\bar O^{K_{(i)}K_{(j)}}_{\chi \psi} &=& \int dw~ \bar \chi^{i_1\cdots i_n} [\nabla_{(i)}^2, \nabla_{(j)}^2] \psi_{i_1\cdots i_n}~.
\label{O-KK}
\eea
We then expand the commutator in the following way,
\bea
[\nabla_{(i)}^2, \nabla_{(j)}^2] &=& \nabla_k[\nabla^{k+i}, \nabla^{l+j}] \nabla_l + \nabla_k \nabla_l[\nabla^{k+i}, \nabla^{l+j}] \cr
&& +[\nabla^{k+i}, \nabla^{l+j}] \nabla_l \nabla_k + \nabla_l[\nabla^{k+i}, \nabla^{l+j}] \nabla_k~.
\label{nabla2-comm}
\eea 
Notice that the terms have been written in such a way that the shifts always appear within the commutators. This can be done as we may write, 
\bea
\nabla^{k+i}\nabla_k = g^{k+i l}\nabla_l \nabla_k = g^{l+i k}\nabla_l \nabla_k = \nabla_k \nabla^{k+i}~,
\eea
without assuming any shift property for a tensor field on which the derivatives are acting. Then we recognize that the following is true for an arbitrary tensor $t^{i_1\cdots i_n}$ which does not necessarily possess the shift property,
\bea
[\nabla^{k+i}, \nabla^{l+j}] t^{i_1\cdots i_n} &=& r^{i_1~~k+i l+j}_{~~i'} t^{i'i_2\cdots i_n} + r^{i_2~~k+i l+j}_{~~i'} t^{i_1i' i_3\cdots i_n} + \cdots ~, \cr
&=& r^{i_1+i+j~~k l}_{~~~~~~~i'} t^{i'i_2\cdots i_n} + r^{i_2+i+j~~k
  l}_{~~~~~~~i'} t^{i_1i' i_3\cdots i_n} + \cdots ~,
\eea
where in the second step we have used the shift property (\ref{shift}) for the Riemann tensor. This shows that the right hand side of (\ref{nabla2-comm}) is symmetric under $i\leftrightarrow j$. However, the left hand side is manifestly anti-symmetric. Therefore, it must vanish.

Let us now consider the linear term $\bar L^{K_{(i)}K_{(j)}}_{\chi
  \psi}$. Following similar argument as above many of the terms can be
shown to vanish by using integration by parts and the shift property
of $\omega_{i_1\cdots i_n j_1\cdots j_n k}$ and its covariant
derivatives. The terms that are problematic are as follows, 
\bea
\bar L^{K_{(i)}K_{(j)}}_{\chi \psi} &=& \int dw~\lt[\nabla^{k+i}\bar \chi^{i_1\cdots i_n} \nabla^{l+j} \omega_{i_1\cdots i_n j_1 \cdots j_n l} \nabla_k \psi^{j_1\cdots j_n} \rt. \cr
&& \lt. + 2\nabla^{k+i} \bar \chi^{i_1\cdots i_n} \omega_{i_1\cdots i_nj_1\cdots j_nl}\nabla_k \nabla^{l+j}\psi^{j_1\cdots j_n} - \nabla_l \bar \chi^{i_1\cdots i_n} \nabla^{k+i} \omega_{i_1\cdots i_n j_1 \cdots j_n k} \nabla^{l+j} \psi^{j_1\cdots j_n}\rt. \cr
&&\lt. -2\nabla_l \nabla^{k+i} \bar \chi^{i_1\cdot i_n} \omega_{i_1\cdots i_n j_1 \cdots j_n k} \nabla^{l+j}\psi^{j_1\cdots j_n}\rt] - i\leftrightarrow j~.
\eea 
Notice that each of the above terms is such that the shift $i$ can not be moved to the position of the shift $j$ and vice versa without using the shift property of the wavefunctions. Therefore, such terms are not individually symmetric under $i\leftrightarrow j$. However, by using integrations by parts the above terms can be rearranged to give the following form,
\bea
\bar L^{K_{(i)}K_{(j)}}_{\chi \psi} &=& \int dw~ \lt[-\nabla^{l+j}\nabla^{k+i} \bar \chi^{i_1\cdots i_n} (\omega_{i_1\cdots i_n j_1\cdots j_n l} \nabla_k +k\leftrightarrow l)\psi^{j_1\cdots j_n} \rt. \cr
&& \lt. +(\nabla_k \bar \chi^{i_1\cdots i_n} \omega_{i_1\cdots i_nj_1\cdots j_nl} + k\leftrightarrow l) \nabla^{k+i} \nabla^{l+j} \psi^{j_1\cdots j_n} \rt. \cr
&& \lt. + [\nabla^{k+i}, \nabla^{l+j}]\bar \chi^{i_1\cdots i_n} \omega_{i_1\cdots i_n j_1\cdots j_n k} \nabla_l \psi^{j_1\cdots j_n} \rt. \cr
&& \lt.+ \nabla_k \bar \chi^{i_1\cdot i_n} \omega_{i_1\cdots i_n j_1\cdots j_n l} [\nabla^{k+i}, \nabla^{l+j}] \psi^{j_1\cdots j_n}\rt] - i\leftrightarrow j~.
\eea
The last two terms are symmetric in $i$ and $j$ because of the reason discussed earlier. Therefore, they drop out in the final result when we anti-symmetrize. Each of the first two terms has a factor symmetric under $k\leftrightarrow l$. Because of this such a term, combined with the counterpart with $i$ and $j$ interchanged, finally contains the commutator of the {\it shifted covariant derivatives} $[\nabla^{k+i}, \nabla^{l+j}]$. Therefore these terms also vanish following the previous argument. Similar arguments can be used to show that all the other relevant terms vanish leading to the result (\ref{T-KK}).

\vspace{.1in}
\noindent
\underline{\large $T^{Z_{(i)}Z_{(j)}}$} \\
Using (\ref{chi-ZL-tensor}, \ref{tensor-ZL-psi}) and manipulating various
terms we first write the contributions to ${}_{\chi}\la\hat Z^L_{(i)}\hat
Z^L_{(j)}\ra_{\psi}$ in the following forms,
\bea
O^{Z^L_{(i)}Z^L_{(j)}}_{\chi \psi} &=& \int dw~ \lt[\nabla_l \nabla_k
\bar \chi^{i_1\cdots i_n} a^{k+i} a^{l+j} \psi_{i_1\cdots i_n} +
\nabla_k \bar \chi^{i_1\cdots i_n} \nabla_l a^{k+i} a^{l+j} \psi_{i_1\cdots
  i_n}\rt]~,
\label{O-ZLZL} \\
L^{Z^L_{(i)}Z^L_{(j)}}_{\chi \psi} &=& \int dw~ \lt[ \lt(\nabla_k
\bar \chi^{i_1\cdots i_n} a^{k+i} \omega_{i_1\cdots i_n j_1\cdots j_nl}
a^{l+j}\psi^{j_1\cdots j_n} + i\leftrightarrow j\rt) \rt. \cr
&& \lt. +\bar \chi^{i_1\cdots i_n} \nabla_l \omega_{i_1\cdots i_n j_1\cdots j_n k}a^{k+i} a^{l+j} \psi^{j_1\cdots j_n} +
\bar \chi^{i_1\cdots i_n} \omega_{i_1\cdots i_n j_1\cdots j_n k} \nabla_l a^{k+i} a^{l+j} \psi^{j_1\cdots j_n} \rt]~, \cr && \label{L-ZLZL} \\
Q^{Z^L_{(i)}Z^L_{(j)}}_{\chi \psi} &=& \int dw~ \bar \chi^{i_1\cdots i_n}
\omega_{i_1\cdots i_n k_1\cdots k_n k} \omega^{k_1\cdots
  k_n}_{~~~~~~j_1\cdots j_n l} a^{k+i} a^{l+j} \psi^{j_1\cdots j_n} ~.
\label{Q-ZLZL}
\eea
Anti-symmetrizing the above results in $i$ and $j$ leads to the
following contributions to the commutator,
\bea
\bar O^{Z^L_{(i)}Z^L_{(j)}}_{\chi \psi} &=& \int dw~ \lt[ [\nabla_l,
\nabla_k] \bar \chi^{i_1\cdots i_n}a^{k+i} a^{l+j} \psi_{i_1\cdots i_n}
\rt. \cr
&& \lt. + i\sum_k \lt\{(k+i) \nabla_k \bar \chi^{i_1\cdots i_n}
a^{k+i+j}\psi_{i_1\cdots i_n} - i\leftrightarrow j \rt\}\rt] ~, \cr
&=& \int dw~ \lt[\bar \chi^{i_1\cdots i_n} r_{i_1\cdots i_n j_1\cdots j_n
  kl} a^{k+i} a^{l+j} \psi^{j_1\cdots j_n} + i(i-j) \nabla_k
\bar \chi^{i_1\cdots i_n} a^{k+i+j} \psi_{i_1\cdots i_n} \rt]~, \cr && \label{Obar-ZLZL}\\
\bar L^{Z^L_{(i)}Z^L_{(j)}}_{\chi \psi} &=& \int dw~ \lt[
\bar \chi^{i_1\cdots i_n}\lt(\nabla_l\omega_{i_1\cdots i_n j_1\cdots j_n k} - \nabla_k \omega_{i_1\cdots i_n j_1\cdots j_n l} \rt) a^{k+i}
a^{l+j} \psi^{j_1\cdots j_n} \rt. \cr
&& \lt. + i\sum_k \lt\{ (k+i) \bar \chi^{i_1\cdots i_n}
\omega_{i_1\cdots i_n j_1 \cdots j_n k} a^{k+i+j} \psi^{j_1\cdots j_n} - i\leftrightarrow j \rt\} \rt]~,\cr
&=& \int dw~ \lt[-\bar \chi^{i_1\cdots i_n} \lt(r_{i_1\cdots i_n j_1\cdots
  j_n kl} + d_{i_1\cdots i_n j_1\cdots j_n kl} \rt) a^{k+i} a^{l+j}
\psi^{j_1\cdots j_n}\rt. \cr
&& \lt. + i(i-j) \bar \chi^{i_1\cdots i_n}\omega_{i_1\cdots i_n j_1\cdots
  j_n k} a^{k+i+j} \psi^{j_1\cdots j_n} \rt] ~, 
\label{Lbar-ZLZL} \\
\bar Q^{Z^L_{(i)}Z^L_{(j)}}_{\chi \psi} &=& \int dw~ \bar \chi^{i_1\cdots
  i_n} \lt(\omega_{i_1\cdots i_n k_1\cdots k_n k} \omega^{k_1\cdots
  k_n}_{~~~~~~j_1\cdots j_n l} -l \leftrightarrow k \rt) a^{k+i} a^{l+j}
\psi^{j_1\cdots j_n} ~, \cr
&=& \int dw~
\bar \chi^{i_1\cdots i_n} d_{i_1\cdots i_n j_1\cdots j_n kl} a^{k+i}
a^{l+j} \psi^{j_1\cdots j_n}~.
\label{Qbar-ZLZL}
\eea
In the first steps of (\ref{Obar-ZLZL}) and (\ref{Lbar-ZLZL}) we have
evaluated the covariant derivative $\nabla_l a^{k+i}$ appearing in (\ref{O-ZLZL}) and (\ref{L-ZLZL}) using (\ref{del-a}). The connection terms coming from these covariant derivatives do not contribute because of the anti-symmetrization in $i$ and $j$. The first terms in (\ref{Obar-ZLZL}) and (\ref{Lbar-ZLZL}) and the term in (\ref{Qbar-ZLZL}) are evaluated using (\ref{chi-r-psi}), (\ref{dalpha}) and
(\ref{alpha-square}) respectively. Each of these terms vanishes individually because of the shift property. However, an interesting feature of these results is the following. Even if we do not use the shift property, such terms cancel each other when we add all the above contributions to get the relevant commutator. We will see as we proceed that this sort of cancellations will be very crucial in computing certain other terms of the DWV algebra. Using (\ref{chi-ZL-psi}) the final result
is written in the following form,
\bea
T^{Z^L_{(i)}Z^L_{(j)}}&=&
-(i-j) {}_{\chi}\la \hat Z^L_{(i+j)} \ra_{\psi}~.
\label{T-ZLZL}
\eea
A similar calculation shows,
\bea
T^{Z^R_{(i)}Z^R_{(j)}} &=&
-(i-j) {}_{\chi}\la \hat Z^R_{(i+j)} \ra_{\psi}~.
\label{T-ZRZR}
\eea
To calculate the last term in (\ref{Z-Z-comm}) we manipulate the
terms in ${}_{\chi}\la \hat Z^L_{(i)} \hat Z^R_{(j)} \ra_{\psi}$ and
${}_{\chi}\la \hat Z^R_{(j)} \hat Z^L_{(i)} \ra_{\psi}$ to arrive at
the following results for the terms in $T^{Z^L_{(i)}Z^R_{(j)}}$ at various orders,
\bea
\bar O^{Z^L_{(i)}Z^R_{(j)}}_{\chi \psi} &=& \int dw~ \lt[\bar \chi^{i_1\cdots
  i_n} a^{l+j} [\nabla_k, \nabla_l] (a^{k+i}\psi_{i_1\cdots i_n} )
+ \nabla_k \bar \chi^{i_1\cdots i_n}a^{l+j} \nabla_l a^{k+i}
\psi_{i_1\cdots i_n} \rt. \cr
&& \lt. + \bar \chi^{i_1\cdots i_n} \nabla_k a^{l+j} a^{k+i}
\nabla_l \psi_{i_1\cdots i_n} + \bar \chi^{i_1\cdots i_n} \nabla_k a^{l+j}
\nabla_l a^{k+i} \psi_{i_1\cdots i_n}\rt] ~, \cr
&=& \int dw~ \lt[\bar \chi^{i_1\cdots i_n} a^{k+i} r_{kl} a^{l+j}
\psi_{i_1\cdots i_n} + \bar \chi^{i_1\cdots i_n} a^{k+i} r_{i_1\cdots i_n
  j_1\cdots j_n kl} a^{l+j}\psi^{j_1\cdots j_n} \rt. \cr
&& + \bar \chi^{i_1\cdots i_n} \nabla_k a^{l+j}
\nabla_l a^{k+i} \psi_{i_1\cdots i_n} + \nabla_k \bar \chi^{i_1\cdots i_n}a^{l+j} \nabla_l a^{k+i}
\psi_{i_1\cdots i_n}  \cr
&& \lt. + \bar \chi^{i_1\cdots i_n} \nabla_k a^{l+j} a^{k+i}
\nabla_l \psi_{i_1\cdots i_n} \rt] ~, \label{Obar-ZLZR} \\
\bar L^{Z^L_{(i)}Z^R_{(j)}}_{\chi \psi} &=& -\int dw~
\bar \chi^{i_1\cdots i_n} \lt[ a^{k+i} a^{l+j} \lt( \nabla_k
\omega_{i_1\cdots i_n j_1\cdots j_n l} -k\leftrightarrow l\rt)
\rt. \cr
&& \lt. + \lt(a^{k+i}\nabla_k
a^{l+j} \omega_{i_1\cdots i_n j_1\cdots j_n l} - i\leftrightarrow j
\rt) \rt]\psi^{j_1\cdots j_n} ~, \cr
&=& \int dw~ \bar \chi^{i_1\cdots i_n} \lt[ -a^{k+i} a^{l+j} \lt(
r_{i_1\cdots i_n j_1\cdots j_n kl} + d_{i_1\cdots i_n j_1 \cdots j_n
  kl}\rt) \rt. \cr
&& \lt. + i(i-j) a^{k+i+j} \omega_{i_1\cdots i_n j_1\cdots j_n k}
\rt]\psi^{j_1\cdots j_n} ~, \label{Lbar-ZLZR} \\
\bar Q^{Z^L_{(i)}Z^R_{(j)}}_{\chi \psi} &=& \int dw~ \bar \chi^{i_1\cdots
  i_n} \lt(\omega_{i_1\cdots i_n k_1 \dots k_n k} \omega^{k_1\cdots
  k_n}_{~~~~~j_1\cdots j_n l} - k\leftrightarrow l \rt) a^{k+i}
a^{l+j} \psi^{j_1\cdots j_n} ~, \cr
&=& \int dw~ \bar \chi^{i_1\cdots i_n} d_{i_1\cdots i_n j_1\cdots j_n kl}
a^{k+i} a^{l+j} \psi^{j_1\cdots j_n}~.
\eea
In this case also the terms involving the Riemann tensor and $d_{i_1\cdots i_n j_1\cdots j_nkl}$ vanish individually due to the shift property. They also cancel each other when we add all the above contributions. Finally, the relevant commutator is given by,
\bea
T^{Z^L_{(i)}Z^R_{(j)}} &=& (-i)^2 \int dw~ \lt[
\bar \chi^{i_1\cdots i_n} a^{k+i}  r_{kl} a^{l+j}
\psi_{i_1\cdots i_n} + \bar \chi^{i_1\cdots i_n} \nabla_k a^{l+j}
\nabla_l a^{k+i} \psi_{i_1\cdots i_n} \rt. \cr
&& + \nabla_k \bar \chi^{i_1\cdots i_n}a^{l+j} \nabla_l a^{k+i}
\psi_{i_1\cdots i_n} + \bar \chi^{i_1\cdots i_n} \nabla_k a^{l+j} a^{k+i}
\nabla_l \psi_{i_1\cdots i_n} \cr
&& \lt. +i(i-j) \bar \chi^{i_1\cdots i_n} a^{k+i+j} \omega_{i_1\cdots i_n
  j_1\cdots j_n k} \psi^{j_1\cdots j_n} \rt]~.
\label{T-ZLZR}
\eea
As indicated in eq.(\ref{Z-Z-comm}), we need to anti-symmetrize the
above quantity in $i$ and $j$. The first two terms drop out as they
are symmetric. The final result is given by,
\bea
T^{Z^L_{(i)}Z^R_{(j)}} - i\leftrightarrow j &=& -(i-j)
{}_{\chi}\la \hat Z_{(i+j)}\ra_{\psi}~.
\label{T-ZLZR-anti}
\eea
Substituting (\ref{T-ZLZL}), (\ref{T-ZRZR}) and (\ref{T-ZLZR}) into
(\ref{Z-Z-comm}) one gets,
\bea
T^{Z_{(i)}Z_{(j)}} &=& -2(i-j) {}_{\chi}\la \hat Z_{(i+j)}\ra_{\psi}~.
\label{T-ZZ}
\eea

\vspace{.1in}
\noindent
{\large \underline{$T^{K_{(i)}Z_{(j)}}$}}\\
This is the most important contribution as the anomaly term arises
here. After a somewhat lengthy calculation we arrive at the following
expressions,
\bea
\bar O^{K_{(i)}Z^L_{(j)}}_{\chi \psi} &=&  \int dw~ \lt[\nabla^{k+i}
\bar \chi^{i_1\cdots i_n} r_{kl} a^{l+j} \psi_{i_1\cdots i_n}
\rt. \cr
&& + \nabla^{l+i}\nabla_k \bar \chi^{i_1\cdots i_n} \nabla_l a^{k+j}
\psi_{i_1\cdots i_n} - \nabla_k \bar \chi^{i_1\cdots i_n} \nabla^{l+i} a^{k+j}
\nabla_l\psi_{i_1\cdots i_n}  \cr
&& -(\nabla^{k+i}\bar \chi^{i_1\cdots i_n} r_{i_1\cdots i_n j_1\cdots
  j_n kl} a^{l+j} \psi^{j_1\cdots j_n} + \bar \chi \leftrightarrow \psi) \cr
&&\lt.
+ \bar \chi^{i_1\cdots i_n}r_{i_1\cdots i_n j_1 \cdots j_n kl}
\nabla^{k+i} a^{l+j} \psi^{j_1\cdots j_n} \rt] ~,
\label{Obar-KZL} \\
\bar L^{K_{(i)}Z^L_{(j)}}_{\chi \psi} &=& -\int dw~
\lt[(\bar \chi^{i_1\cdots i_n} \omega_{i_1\cdots i_n j_1\cdots j_n l}\nabla_k
\psi^{j_1\cdots j_n} + \bar \chi \leftrightarrow \psi) (\nabla^{k+i}
a^{l+j} + k\leftrightarrow l) \rt. \cr
&& - \{ \nabla^{k+i} \bar \chi^{i_1\cdots i_n}  (r_{i_1\cdots i_n j_1\cdots j_n kl} + d_{i_1\cdots i_n j_1\cdots j_n kl}) a^{l+j} \psi^{j_1\cdots j_n} + \bar \chi \leftrightarrow  \psi \}  \cr
&& + 2\bar \chi^{i_1\cdots i_n} r_{i_1\cdots i_n
  k_1\cdots k_n kl} \omega^{k_1\cdots k_n~~~~~~k+i}_{~~~~~~j_1\cdots
  j_n} a^{l+j} \psi^{j_1\cdots j_n} \cr
&& + \bar \chi^{i_1\cdots i_n} \{\nabla_l(\omega_{i_1\cdots i_n j_1\cdots
  j_n}^{~~~~~~~~~~~~k+i} \nabla_k a^{l+j})  -\nabla_l \omega_{i_1\cdots i_n
j_1\cdots j_n k}\nabla^{k+i} a^{l+j} \cr
&& \lt. - [\nabla_k, \nabla_l]\omega_{i_1\cdots i_n
  j_1\cdots j_n}^{~~~~~~~~~~~~k+i} a^{l+j}  \} \psi^{j_1\cdots j_n} \rt] ~,
\label{Lbar-KZL} \\
\bar Q^{K_{(i)}Z^L_{(j)}}_{\chi \psi} &=& \int dw~
\lt[-\nabla^{k+i} \bar \chi^{i_1\cdots i_n}d_{i_1\cdots i_n j_1\cdots j_n kl} a^{l+j}
\psi^{j_1\cdots j_n} \rt. \cr
&& + \bar \chi^{i_1\cdots i_n} (r_{i_1\cdots i_n  k_1\cdots k_n kl} +d_{i_1\cdots i_n  k_1\cdots k_n kl})\omega^{k_1\cdots k_n~~~~~~k+i}_{~~~~~~j_1\cdots j_n}
a^{l+j} \psi^{j_1\cdots j_n} \cr
&& \lt. + \bar \chi^{i_1\cdots i_n}\omega_{i_1\cdots i_n k_1\cdots k_n
  k}\omega^{k_1\cdots k_n}_{~~~~~~j_1\cdots j_n l} \nabla^{k+i} a^{l+j}
\psi^{j_1\cdots j_n} + \bar \chi \leftrightarrow \psi \rt] ~,
\label{Qbar-KZL} \\
\bar C^{K_{(i)}Z^L_{(j)}}_{\chi \psi} &=& -\int dw~
\lt[\bar \chi^{i_1\cdots i_n} \omega_{i_1\cdots i_n k_1\cdots k_n k}
\omega^{k_1\cdots k_n~~~~~k+i}_{~~~~~~l_1\cdots l_n}\omega^{l_1\cdots
  l_n}_{~~~~~j_1\cdots j_n l} a^{l+j} \psi^{j_1\cdots j_n} + \bar \chi
\leftrightarrow \psi \rt]~, \cr
&=& -\int dw~ \lt[\bar \chi^{i_1\cdots i_n}d_{i_1\cdots i_n k_1 \cdots k_n
  kl} \omega^{k_1\cdots k_n~~~~~~k+i}_{~~~~~~j_1\cdots j_n} a^{l+j}
\psi^{j_1\cdots j_n} + \bar \chi \leftrightarrow \psi \rt] ~.
\label{Cbar-KZL}
\eea
To get to the second step of (\ref{Cbar-KZL}) we have used
(\ref{alpha-square}) twice to push
$\omega^{l_1\cdots  l_n}_{~~~~~~j_1\cdots j_n l}$ to the left.
In order to derive the expected result for $T^{K_{(i)}Z_{(j)}}$ we need cancellation of many terms to take place. This happens in two steps - some terms get
canceled when we add all the above contributions to find
$T^{K_{(i)}Z^L_{(j)}}$. Another set of cancellations takes place when we add the latter to $T^{K_{(i)}Z^R_{(j)}}$ to get the final result for $T^{K_{(i)}Z_{(j)}}$. To arrive at our final results, however, it will be easier for us to go in the reverse order. One can be easily convinced that various contributions to $T^{K_{(i)}Z^R_{(j)}}$ can be obtained by interchanging $\bar \chi$ and $\psi$ in the expressions for the corresponding contributions to $T^{K_{(i)}Z^L_{(j)}}$. Therefore to get the terms in $T^{K_{(i)}Z_{(j)}}$ at various orders we simply symmetrize the
expressions in (\ref{Obar-KZL}, \ref{Lbar-KZL}, \ref{Qbar-KZL}) and
(\ref{Cbar-KZL}) under $\bar \chi \leftrightarrow \psi$,
\bea
\bar O^{K_{(i)}Z_{(j)}}_{\chi \psi} &=& \int dw~
\lt[\nabla^{k+i}\bar \chi^{i_1\cdots i_n}r_{kl} a^{l+j}\psi_{i_1\cdots
  i_n} + \nabla^{l+i}\nabla_k \bar \chi^{i_1\cdots i_n} \nabla_l a^{k+j}
\psi_{i_1\cdots i_n} \rt. \cr
&&  - \nabla_k \bar \chi^{i_1\cdots i_n} \nabla^{l+i} a^{k+j}
\nabla_l\psi_{i_1\cdots i_n} - 2\nabla^{k+i} \bar \chi^{i_1\cdots i_n} r_{i_1\cdots i_n j_1\cdots j_n kl} a^{l+j} \psi^{j_1\cdots j_n} \cr
&& \lt. + \bar \chi \leftrightarrow \psi \rt]~,
\label{Obar-KZ} \\
\bar L^{K_{(i)}Z_{(j)}}_{\chi \psi} &=& \int dw~ \lt[- 2\bar \chi^{i_1\cdots i_n}\omega_{i_1\cdots i_n j_1\cdots j_n l}\nabla_k \psi^{j_1\cdots j_n} (\nabla^{k+i}a^{l+j} + k\leftrightarrow l) \rt. \cr
&& +2\nabla^{k+i} \bar \chi^{i_1\cdots i_n}  (r_{i_1\cdots i_n j_1\cdots j_n kl} + d_{i_1\cdots i_n j_1\cdots j_n kl}) a^{l+j} \psi^{j_1\cdots j_n} \cr
&& \lt. - 2\bar \chi^{i_1\cdots i_n} r_{i_1\cdots i_n k_1\cdots k_n kl} \omega^{k_1\cdots k_n~~~~~~k+i}_{~~~~~~j_1\cdots j_n} a^{l+j} \psi^{j_1\cdots j_n} + \bar \chi \leftrightarrow \psi\rt]~,
\label{Lbar-KZ} \\
\bar Q^{K_{(i)}Z_{(j)}}_{\chi \psi} &=& 2 \bar
Q^{K_{(i)}Z^L_{(j)}}_{\chi \psi}~,
\label{Qbar-KZ} \\
\bar C^{K_{(i)}Z_{(j)}}_{\chi \psi} &=& 2 \bar
C^{K_{(i)}Z^L_{(j)}}_{\chi \psi}~.
\label{Cbar-KZ}
\eea
Notice that in the process of symmetrization the last terms of
(\ref{Obar-KZL}) and (\ref{Lbar-KZL}) dropped out. We now add all the
above contributions to get $T^{K_{(i)}Z_{(j)}}$,
\bea
T^{K_{(i)}Z_{(j)}} &=& (-i)^3 \int dw~ \lt[\nabla^{k+i}\bar \chi^{i_1\cdots i_n}r_{kl} a^{l+j}\psi_{i_1\cdots i_n} + \nabla^{l+i}\nabla_k \bar \chi^{i_1\cdots i_n}  \nabla_l a^{k+j} \psi_{i_1\cdots i_n} \rt. \cr
&&  - \nabla_k \bar \chi^{i_1\cdots i_n} \nabla^{l+i} a^{k+j}
\nabla_l\psi_{i_1\cdots i_n} - 2\bar \chi^{i_1\cdots i_n}\omega_{i_1\cdots i_n j_1\cdots j_n l}\nabla_k \psi^{j_1\cdots j_n} (\nabla^{k+i}a^{l+j} + k\leftrightarrow l) \cr
&& \lt.+ 2\bar \chi^{i_1\cdots i_n}\omega_{i_1\cdots i_n k_1\cdots k_n
  k}\omega^{k_1\cdots k_n}_{~~~~~~j_1\cdots j_n l} \nabla^{k+i} a^{l+j}
\psi^{j_1\cdots j_n} + \bar \chi \leftrightarrow \psi \rt]~, \cr
&=&(-i)^3 \int dw~ \lt[\nabla^{k+i}\bar \chi^{i_1\cdots i_n}r_{kl} a^{l+j}\psi_{i_1\cdots i_n} + \bar \chi^{i_1\cdots i_n} a^{k+j} r_{kl} \nabla^{l+i}\psi_{i_1\cdots i_n} \rt. \cr
&& +(\nabla^{l+i}\nabla_k \bar \chi^{i_1\cdots i_n} \nabla_l a^{k+j}
\psi_{i_1\cdots i_n} + \bar \chi \leftrightarrow \psi) \cr
&&  - \{\nabla_k\bar \chi^{i_1\cdots i_n} \nabla_l\psi_{i_1\cdots i_n}
+ 2(\bar \chi^{i_1\cdots i_n}\omega_{i_1\cdots i_n j_1\cdots j_n l}\nabla_k
\psi^{j_1\cdots j_n} -\nabla_k \bar \chi^{i_1\cdots i_n} \omega_{i_1\cdots i_n j_1\cdots j_n l} \psi^{j_1\cdots j_n}) \cr
&& \lt. - 2\bar \chi^{i_1\cdots i_n}\omega_{i_1\cdots i_n k_1\cdots  k_n k}\omega^{k_1\cdots k_n}_{~~~~~~j_1\cdots j_n l} \psi^{j_1\cdots j_n} \} (\nabla^{k+i} a^{l+j} + k\leftrightarrow l)\rt]~,
\label{T-KZ1}
\eea
where in the last step we have simply rearranged the terms in a different way.
We will now manipulate the terms in the last three lines in certain ways. Let us first consider the terms in the second line.
\bea
\nabla^{l+i}\nabla_k\bar \chi^{i_1\cdots i_n}\nabla_l
a^{k+j}\psi_{i_1\cdots i_n}
&=&i\sum_k(k+j)\nabla^{k+i+j}\nabla_k\bar \chi^{i_1\cdots i_n}
\psi_{i_1\cdots i_n} \cr
&& + a^{l'} \gamma^{k+j}_{l'l}
\nabla^{l+i}\nabla_k\bar \chi^{i_1\cdots i_n}\psi_{i_1\cdots i_n} ~, \cr
&=&i\sum_k(k+j)\nabla^{k+i+j}\nabla_k\bar \chi^{i_1\cdots i_n}
\psi_{i_1\cdots i_n} \cr
&& + {1\over 2} a^{l'} \gamma^{k+i+j}_{l'\tilde k}g^{\tilde k l}
\lt(\{\nabla_l,\nabla_k \} +[\nabla_l,\nabla_k ]\rt)
\bar \chi^{i_1\cdots i_n}\psi_{i_1\cdots i_n} ~, \cr
&=& i\sum_k(k+j)\nabla^{k+i+j}\nabla_k \bar \chi^{i_1\cdots i_n}
\psi_{i_1\cdots i_n} \cr
&& +{1\over 2} a^{l'} (\gamma^{k+i+j}_{l'\tilde k}g^{\tilde k l} +
\gamma^{l+i+j}_{l'\tilde k}g^{\tilde k k}) \nabla_l\nabla_k
\bar \chi^{i_1\cdots i_n}\psi_{i_1\cdots i_n} \cr
&& + {1\over 2} a^{l'} \gamma^{k+i+j}_{l'\tilde k}g^{\tilde k l}
\bar \chi^{i_1\cdots i_n} r_{i_1\cdots i_n j_1\cdots j_n kl}
\psi^{j_1\cdots j_n}~.
\label{sec-term1}
\eea
Following \cite{pm0912} we identify the term inside the round bracket in the second line as $-\del_{l'}g^{k+i+j}$ and then use the identity (\ref{id}). This gives for the whole term in the second line,
\bea
&& -{i\over 2} \sum_{k,l}(k+l+i+j) g^{k+i+j l} \nabla_l \nabla_k \bar \chi^{i_1\cdots i_n}\psi_{i_1\cdots i_n} \cr &=& -{i\over 2} (i+j) \nabla_{(i+j)}^2 \bar \chi^{i_1\cdots i_n} \psi_{i_1\cdots i_n} 
-{i \over 2} \sum_k (k) [\nabla^{k+i+j} \nabla_k + \nabla_k \nabla^{k+i+j}] \bar \chi^{i_1\cdots i_n} \psi_{i_1\cdots i_n} ~.\cr &&
\eea
The sum over $k$ in the second term is not a tensor contraction because of the presence of the factor $(k)$. For the same reason the two terms inside the square bracket are not same. However, because of the anti-symmetry property of $r_{i_1\cdots i_n j_1 \cdots j_n kl}$ in (\ref{r-d-anti-symmetry}) such terms give the same result once we symmetrize between $\bar \chi$ and $\psi$,
\bea
\nabla^{k+i+j} \nabla_k \bar \chi^{i_1\cdots i_n} \psi_{i_1\cdots i_n} +\bar \chi \leftrightarrow \psi 
 = \nabla_k \nabla^{k+i+j} \bar \chi^{i_1\cdots i_n} \psi_{i_1\cdots i_n} +\bar \chi \leftrightarrow \psi ~,
\eea
where the index $k$ is not summed over. Because of the same reason the last term in (\ref{sec-term1}) drops out when we symmetrize between $\bar \chi$ and $\psi$. The final result for the second line in (\ref{T-KZ1}) is given by,
\bea
\nabla^{l+i}\nabla_k \bar \chi^{i_1\cdots i_n}\nabla_l a^{k+j}\psi_{i_1\cdots i_n} + \bar \chi \leftrightarrow \psi &=& {i\over 2} (j-i) (\nabla_{(i+j)}^2 \bar \chi^{i_1\cdots i_n}\psi_{i_1\cdots i_n} + \bar \chi^{i_1\cdots i_n}\nabla_{(i+j)}^2 \psi_{i_1\cdots i_n}) ~, \cr &&
\label{sec-term}
\eea
To evaluate the terms in the last two lines of (\ref{T-KZ1}) we first derive,
\bea
\nabla^{k+i} a^{l+j} + k \leftrightarrow l &=& i(j-i) g^{k+i+j l}~,
\label{cov-der-a-symm}
\eea
using the same trick as described below eq.(\ref{sec-term1}). Using (\ref{sec-term}) and (\ref{cov-der-a-symm}) in (\ref{T-KZ1}) one finally finds,
\bea
T^{K_{(i)}Z_{(j)}} &=& -2 (i-j) {}_{\chi}\la\hat
K_{(i+j)}\ra_{\psi} + {}_{\chi}\la \hat \pi^{\star k+i}
r_{kl}(\hat x) a^{l+j}(\hat x) - a^{k+j}(\hat x) r_{kl}(\hat x) \hat
\pi^{l+i} \ra_{\psi}~, \cr &&
\label{T-KZ}
\eea
where we have used eqs.(\ref{chi-K-psi}, \ref{tensorx-pih-psi}).

\vspace{.1in}
\noindent
{\large \underline{The rest}}\\
The rest of the commutators are straightforward to calculate and do not introduce any more complications than in the case of spin-zero representation studied in \cite{pm0912}. The relevant results are as follows,
\bea
T^{K_{(i)}V_{(j)}} - i\leftrightarrow j &=& -2(i-j) {}_{\chi}\la \hat Z_{(i+j)}\ra_{\psi}~,
\label{T-KV-antisymm} \\
T^{Z^L_{(i)}V_{(j)}} &=& T^{Z^R_{(i)}V_{(j)}} = -(i-j) {}_{\chi}\la \hat V_{(i+j)}\ra_{\psi}~.
\label{T-ZLZRV}
\eea

Finally, using the results (\ref{T-KK}, \ref{T-ZZ}, \ref{T-KZ},
\ref{T-KV-antisymm}, \ref{T-ZLZRV}) in equations (\ref{Z-Z-comm},
\ref{K-Z-comm}, \ref{Z-V-comm}) and (\ref{DWValg-parts}) suitably one
finds the algebra as shown in (\ref{DWValg}, \ref{DWVanomaly}).

\end{document}